\documentclass[preprint]{aastex631}

\usepackage{graphicx}
\usepackage{natbib}
\usepackage{amssymb,amsmath}

\usepackage{multirow}
\usepackage{upgreek}
\shorttitle{}
\shortauthors{Stores et al.}
\graphicspath{{./}{figures/}}

\begin{document}

\title{Spectral and Imaging Diagnostics of Spatially-Extended Turbulent Electron Acceleration and Transport in Solar Flares}

\author[0000-0002-6060-8048]{Morgan Stores}
\affiliation{Department of Mathematics, Physics \& Electrical Engineering, Northumbria University,  \\
Newcastle upon Tyne, UK\\
NE1 8ST}

\author[0000-0001-6583-1989]{Natasha L. S. Jeffrey}
\affiliation{Department of Mathematics, Physics \& Electrical Engineering, Northumbria University,  \\
Newcastle upon Tyne, UK\\
NE1 8ST}

\author[0000-0002-7863-624X]{James A. McLaughlin}
\affiliation{Department of Mathematics, Physics \& Electrical Engineering, Northumbria University,  \\
Newcastle upon Tyne, UK\\
NE1 8ST}

\begin{abstract}
Solar flares are efficient particle accelerators with a large fraction of released magnetic energy ($10-50$\%) converted into energetic particles such as hard X-ray producing electrons. This energy transfer process is not well constrained, with competing theories regarding the acceleration mechanism(s), including MHD turbulence. 
We perform a detailed parameter study examining how various properties of the acceleration region, including its spatial extent and the spatial distribution of turbulence, affect the observed electron properties, such as those routinely determined from X-ray imaging and spectroscopy.
Here, a time-independent Fokker-Planck equation is used to describe the acceleration and transport of flare electrons through a coronal plasma of finite temperature. 
Motivated by recent non-thermal line broadening observations that suggested extended regions of turbulence in coronal loops, an extended turbulent acceleration region is incorporated into the model.
We produce outputs for the density weighted electron flux, a quantity directly related to observed X-rays, modelled in energy and space from the corona to chromosphere.
We find that by combining several spectral and imaging diagnostics (such as spectral index differences or ratios, energy or spatial-dependent flux ratios, and electron depths into the chromosphere) the acceleration properties, including the timescale and velocity dependence, can be constrained alongside the spatial properties. Our diagnostics provide a foundation for constraining the properties of acceleration in an individual flare from X-ray imaging spectroscopy alone, and can be applied to past, current and future observations including those from RHESSI and Solar Orbiter.
\end{abstract}

\section{Introduction}
Solar flares are one environment in which the Sun is able to accelerate particles. Using stored magnetic energy released via magnetic reconnection \citep[e.g., ][]{parker1957sweet,1958IAUS....6..123S,2000mare.book.....P}, particles can be accelerated to keV and MeV energies \citep{2008LRSP....5....1B}. However, the processes that ultimately transfer energy to particles, and the primary location(s) of acceleration are not well constrained. 
During most flares we observe deka-keV energetic electrons trapped on newly formed closed magnetic field lines, inferred by their bremsstrahlung X-ray emission \citep[e.g., ][]{holman2011implications}. Thus, hard X-ray (HXR) emission has been a vital tool in determining the properties of flare-accelerated electrons \citep[e.g.,][]{kontar2011deducing} since the first pioneering observations of the 1950s and 1960s \cite[e.g., ][]{1959JGR....64..697P,1969ApJ...158L.159F}.

\cite{1981ApJ...246L.155H} identified HXR sources at the ends of coronal loops, now routinely observed and referred to as HXR footpoints, while \cite{1994Natur.371..495M} uncovered a HXR source in the coronal looptop alongside the footpoints. The imaging spectroscopy abilities of the (Reuven) Ramaty High-Energy Solar Spectroscopic Imager  (RHESSI; \citealt{lin2003reuven}) and now the Spectrometer/Telescope for Imaging X-rays (STIX; \citealt{2020A&A...642A..15K}) onboard Solar Orbiter (SolO; \citealt{2020A&A...642A...1M}) provide spatially resolved X-ray spectra in both the coronal looptop and HXR footpoint sources \citep[e.g., ][]{2003ApJ...595L.107E,2013A&A...551A.135S,2022SoPh..297...93M}.
These observations provide validation to the `thick target' model \citep[e.g.,][]{brown1971deduction} in which electrons are accelerated in the corona and propagate down into the dense chromosphere losing their energy via Coulomb collisions and producing X-ray emission mainly via electron-ion bremsstrahlung. The energy and rate of energetic electrons can influence the size and shape of the X-ray emitting region, with higher energy electrons propagating further into the chromosphere \citep[e.g., ][]{2002SoPh..210..383A, 2010ApJ...717..250K}. Non-thermal electron transport is often described by a ‘cold-thick target’ model (CTTM), in which electron energy $E >> T$, where $T$ is the temperature of the ambient plasma.
Whilst able to describe non-thermal electrons when they reach cooler chromospheric layers, this model is not successful in accounting for high coronal temperatures found during a flare. The `low energy cutoff' problem is a well known issue with the CTTM, where low-energy electrons, and hence the total electron power, cannot be constrained by X-ray spectroscopy. Consideration of thermalization led to the warm target model (WTM) \citep{2015ApJ...809...35K,2019ApJ...871..225K}, where coronal plasma properties help to constrain the properties of non-thermal electrons, overcoming the low energy cutoff problem.

In recent years, the dynamic nature of flares has favoured stochastic `Fermi-type' acceleration models (e.g., magnetohydrodynamic (MHD) plasma turbulence, plasma waves) over large-scale direct-current mechanisms \citep[e.g.,][]{larosa1993mechanism,petrosian2012stochastic}. Further, many models predict turbulence is vital in transferring energy from large-scale magnetic fields to the small-scale particle regime \citep[e.g., ][]{larosa1993mechanism}, backed by abundant observational evidence of turbulence in flares \citep[e.g.,][]{kontar2017turbulent}. Macroscopic random plasma motions in active regions and flares are routinely detected via the presence of `non-thermal broadening', whereby the width of an optically-thin spectral line is greater than expected from ion thermal motions alone \citep[e.g.,][]{doschek1980high,antonucci1995temperature}. Non-thermal broadening is often greatest in regions that coincide to the magnetic looptops \citep[e.g.,][]{doschek2014plasma}, where it is expected (in a standard model) that the primary energy release and the bulk of particle energization occur. \citet{2020ApJ...900..192F} observe turbulence surrounding the plasma sheet above the looptops, possibly due to the presence of the tearing mode instability and the creation of magnetic islands.
Moreover, \cite{kontar2017turbulent} examined the kinetic energy associated with turbulent non-thermal broadening alongside the power associated with energetic electrons, inferring a turbulent dissipation time of 1-10~s.
Such times are consistent with MHD turbulence models \citep[e.g, ][]{1995ApJ...438..763G}. A more detailed study of the same flare by \cite{2021ApJ...923...40S} suggested turbulence has a more complex temporal and spatial structure than previously assumed, with non-thermal broadening present throughout the coronal loops, not just at the looptops. Further, \cite{2021ApJ...923...40S} introduced turbulent kinetic energy maps showing the availability of turbulent kinetic energy throughout the loops and cusp, and identified important spatial inhomogeneities in the plasma motions leading to turbulence.

Previous studies have also explored the connection between X-ray looptop and footpoint sources. \cite{Petrosian2002} studied 18 flares using the Yohkoh/HXR Telescope \citep{1991SoPh..136...17K} and determined on average that the X-ray spectral index $\gamma$ was greater in the looptops than the footpoints, by $\approx1$. However, the data here were limited by the resolution of Yohkoh. \cite{2006A&A...456..751B} determined a spectral index difference closer to 2 in most flares ($1.8$), which is expected in a simple scenario where electrons stream from a thin-target corona to a thick-target chromosphere, but some flares showed differences $\pm 2$.  Further, \cite{2013A&A...551A.135S} showed that a portion of the accelerated electron population must be trapped in the looptop to account for the higher acceleration electron rates in the looptop than the footpoints. Accelerated electrons may experience magnetic trapping or pitch-angle scattering preventing the electrons from leaving the coronal looptop. Within turbulent regions, charged particles may become trapped by turbulent scattering due to magnetic fluctuations \citep{musset2018diffusive}. Thus, the presence of turbulence in flares is intimately linked with both electron acceleration and transport.

The insightful introduction of a spatially dependent turbulent acceleration diffusion coefficient in \cite{stackhouse2018spatially}, shows the importance of accounting for a spatial variation in turbulence; producing a softer spectrum than the spatially-averaged description of electron acceleration and transport given in the leaky box model \citep{2013ApJ...777...33C}. Following the observational results of \cite{2021ApJ...923...40S} and the preliminary work of \cite{stackhouse2018spatially}, we perform a detailed parameter study examining how the presence and varying properties of a spatially-extended turbulent region in the corona changes the spectral and spatial (imaging) properties of observed flare-accelerated electrons deduced from X-ray spectroscopy and imaging. \S \ref{sec:method} provides an overview of the electron acceleration and transport model, \S \ref{results} discusses the main results, and \S \ref{sec:summary} summarises the study and its application to HXR data.

\section{Combined acceleration and transport modelling of energetic electrons}
\label{sec:method}
\begin{figure*}
    \centering
    \includegraphics[scale = 0.37]{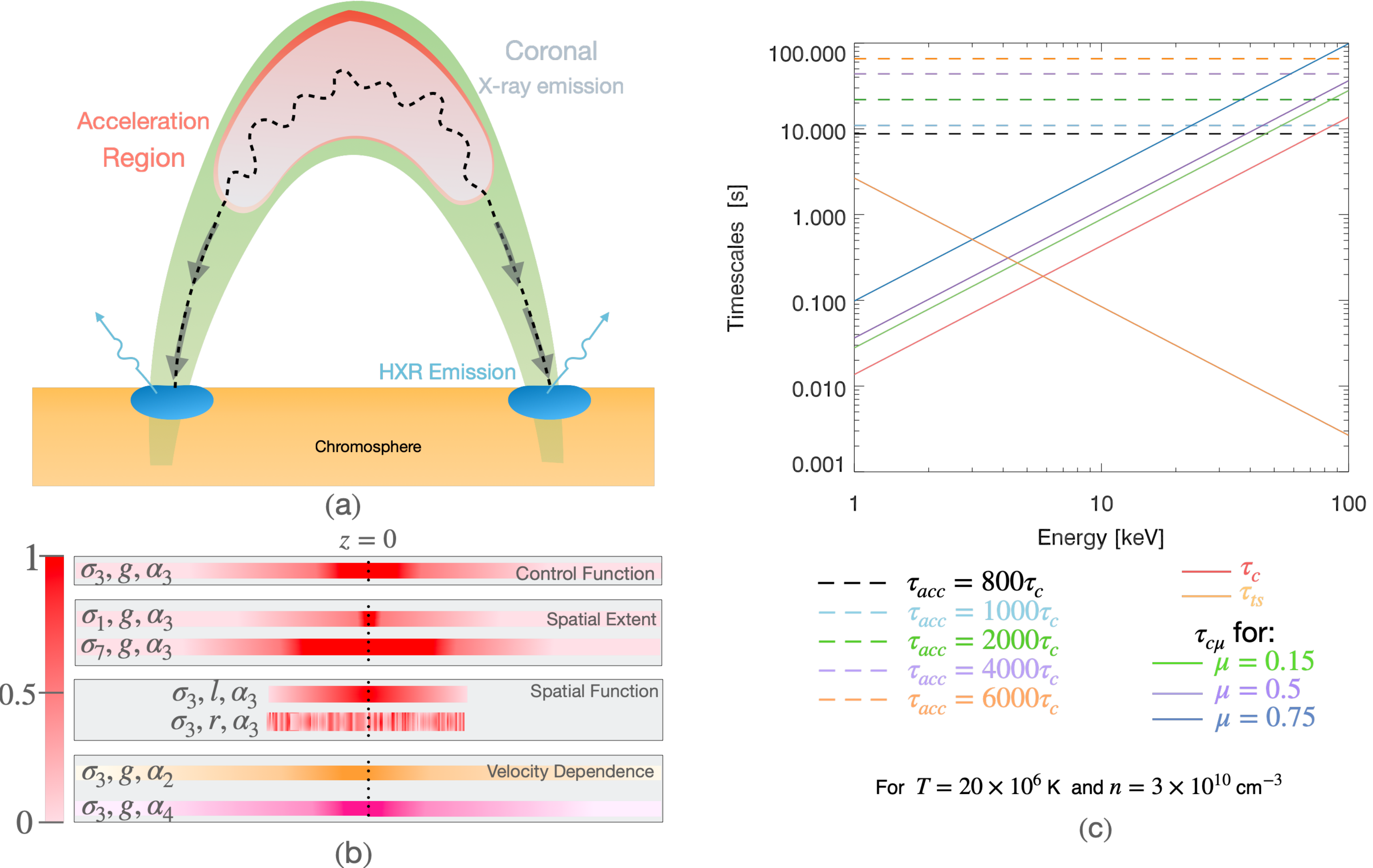}
    \caption{(a) Cartoon of an extended acceleration region in a flare loop, showing X-ray emission from a looptop source and HXR emission from chromospheric footpoints. Black arrows indicate direction of electron transport down the coronal loop. 
    (b) Diagram listing the simulated accelerated regions. The acceleration region is depicted by a colored rectangle, centred at the coronal looptop ($z = 0\arcsec$) indicated by the dashed black line. The strength (greater diffusion) of the acceleration region is demonstrated qualitatively by the color gradient. Each row highlighted by the gray box shows the following: Row 1 - the control simulation referred to as $\sigma_3,\rm{g},\alpha_3$, Row 2 - simulations with different spatial extents ($\sigma_1,\rm{g},\alpha_3$ and $\sigma_7,\rm{g},\alpha_3$), Row 3 - simulations with different spatial functions ($\sigma_3,\textit{l},\alpha_3$ and $\sigma_3,\textit{r},\alpha_3$), and Row 4 - simulations with different velocity dependencies, ($\sigma_3,\rm{g},\alpha_2$ and $\sigma_3,\rm{g},\alpha_4$), shown by orange and pink acceleration regions, respectively. 
    (c) Relevant simulation timescales $\uptau$ versus electron energy $E$. The colored dashed lines indicate different acceleration timescales $\uptau_{acc}$ where $A_{acc} = 800,1000,2000,4000,6000,$ shown in black, blue, green, purple, and orange, respectively. The solid red line shows the collisional deceleration timescale $\uptau_c$. The collisional scattering timescale $\uptau_{c\mu}$ for $\mu = 0.15, 0.5$ and 0.75 are shown by the solid green, purple, and blue lines, respectively. An energy-dependent turbulent scattering timescale (Equation \ref{tsmfp}) $\uptau_{ts}$ is denoted by the orange solid line.}
    \label{fig:1}
\end{figure*}

\subsection{Governing Fokker-Planck equation}
A time-independent Fokker-Planck equation \citep[e.g][]{holman2011implications,battaglia2012numerical,kontar2015collisional,jeffrey2019role} is used to describe the evolution of an electron flux spectrum $F(E,z,\mu)$ [electrons cm$^{-2}$ s$^{-1}$ keV$^{-1}$], which is a function of field-aligned coordinate $z$ [cm], energy $E$ [keV] and cosine of the pitch-angle ($\beta$) to the guiding magnetic field $\mu=\cos\beta$. The time-independent Fokker-Plank equation is useful for studying electron transport when the transport time from the corona to the lower atmosphere is shorter than the observational time \citep{2014ApJ...787...86J}\footnote{X-ray imaging times are usually tens of seconds to minutes to provide enough counts in the studied energy range.}.

A Fokker-Planck equation describing both stochastic acceleration and scattering, and collisional transport through a warm coronal plasma of temperature $T$ [K], number density $n$ [cm$^{-3}$] and length $L$ [cm] may be written as,

\begin{equation}
\begin{split}
        \mu \frac{\partial F}{\partial z} & =
        \underbrace{\sqrt{2m_e^3} \bigg \{\frac{\partial}{\partial E}\left[ E^{3/2} D(v,z) \frac{\partial }{\partial E} \left( \frac{F}{E}\right )\right]}_\text{turbulent acceleration}\bigg \} \\
        &+ \underbrace{\Gamma m_e^2 \bigg \{ \frac{\partial}{\partial E} \bigg [ G(u[E]) \frac{\partial F}{\partial E} + \frac{G(u[E])}{E}\bigg ( \frac{E}{k_B T} - 1\bigg ) F \bigg ]}_\text{collisional energy losses} \bigg \} \\
        & + \underbrace{\sqrt{\frac{m_{e}}{2E}}\bigg \{\frac{\partial}{\partial \mu}\left[D_{\mu\mu}(\mu,v,z)\frac{\partial F}{\partial \mu}\right]\bigg \}}_\text{turbulent scattering} \\
        & + \underbrace{\frac{\Gamma m_e^2}{8E^2} \bigg \{\frac{\partial}{\partial \mu} \bigg [ (1 - \mu^2) \left [ \text{erf}(u[E]) - G(u[E])\right ]\frac{\partial F }{\partial \mu} \bigg ] \bigg \}}_\text{collisional pitch-angle scattering} \\
        & + S_F(E,z,\mu) 
        \end{split}
        \label{time_independent_Fokker_planck}
\end{equation}
where $v$ is velocity [cm s$^{-1}$], and $\Gamma = 4 \pi e^4$ln$\Lambda n / m_e^2$, for electron charge $e$ [statC], Coulomb logarithm ln$\Lambda$, electron mass $m_e$ [g], and $k_B$ is the Boltzmann constant. $S_{F}$ is the source term further discussed in \S \ref{plasma_conditions}.

The error function erf$(u)$ and the Chandrasekhar function $G(u)$ are given by:
\begin{equation}
    \text{erf}(u) \equiv \frac{2}{\sqrt{\pi}} \int_0^u \text{exp}(-t^2) dt
\end{equation}
and 
\begin{equation}
    G(u) \equiv \frac{\text{erf}(u) - u \, \text{erf}\,'(u)}{2u^2}
\end{equation}
where $u$ is the dimensionless velocity $u = v/(\sqrt{2} v_{th})$, $v_{th} = \sqrt{k_BT/m_e}$ and $\text{erf}\,'(u)= d\text{erf}/du$. Such functions control the lower-energy ($E\approx k_{B}T$) electron interactions ensuring that they become indistinguishable from the background thermal plasma.

$D(v,z)$ is the turbulent acceleration diffusion coefficient used to accelerate electrons out of a thermal plasma and $D_{\mu\mu}(\mu,v,z)$ is the turbulent scattering coefficient. Both are discussed in more detail in \S \ref{sec:turb_acc}.

\subsection{Stochastic differential equations (SDEs)}\label{sdes}
To allow the evolution of an electron distribution to be modelled in space, energy, and pitch-angle to the guiding magnetic field, Equation \ref{time_independent_Fokker_planck} can be solved numerically by its conversion into a set of time-independent stochastic differential equations (SDEs) \citep[e.g.,][]{1986ApOpt..25.3145G,2017SSRv..212..151S}:
\begin{equation}
    z_{j+1} = z_j + \mu_j \Delta s \,\,\,\,\, ,
    \label{eq:sde_z}
\end{equation}
\begin{equation}
    \begin{split}
    &\mu_{j+1} =  \mu_j \\
    & - \left\{  \frac{\Gamma_{\textit{\text{eff}}}\left[\text{erf}(u_j) - G(u_j) \right ] }{4E^2_j} -\sqrt{\frac{2m_e}{E_j}}\frac{ D_{\mu\mu}(\mu,v,z)}{(1-\mu_j^2)}\right \} \mu_j\Delta s \\
    & + \Bigg( \Bigg\{ \frac{(1-\mu^2_j)\Gamma_{eff}\left [ \text{erf}(u_j) - G(u_j) \right ] }{4E^2_j}  \\
    &  +  D_{\mu\mu}(\mu,v,z) \sqrt{\frac{m_e}{2E_j}} \, \Bigg\} \Delta s  \Bigg)^{1/2} W_\mu\\
    \end{split}
    \label{eq:sde_mu}
\end{equation}
and 
\begin{equation}
\begin{split}
    E_{j+1} = &E_j - \frac{\Gamma_{\textit{eff}}}{2E_j}[\text{erf}(u_j)-2u_j\text{erf}'(u_j)] + 3\sqrt{\frac{m_e^3}{2E_j}} D(v,z) \Delta s \\
    & + \sqrt{2m_e^3E_j} \frac{\partial }{\partial E_j} \left[  D(v,z)\right ] \Delta s \\
    &+ \sqrt{2[(2m_e^3E_j)D(v,z) +\Gamma_{\textit{eff}}G(u_j)]\Delta s }\,W_E
\end{split}
    \label{eq:sde_e}
\end{equation}

where $\Gamma_{\textit{eff}} = \Gamma Z_{\textit{eff}} m_e^2$ and $Z_{\textit{eff}}$ is the effective atomic number (set to 1 for electron-electron collisions), $W_\mu$ and $W_E$ are the Wiener functions, and $\Delta s$ is the length step, which is the maximum distance the electron can travel in each simulation iteration. The thermal collisional length is given by $\lambda_c=v_{\rm th}\uptau_c$ where the thermal collisional time $\uptau_c\approx v_{\rm th}^{3}/\Gamma$. $\lambda_c$ can be used to determine an appropriate value of $\Delta s$, since $\Delta s<\lambda_c$. Here, $\Delta s=1\times10^{5}$~cm which is around two orders of magnitude smaller than $\lambda_c$ for our chosen plasma parameters (\ref{plasma_conditions}) and $\sim$ two orders of magnitude smaller than the turbulent scattering length $\lambda_{ts}(E)$ at $E=100$~keV (see \S \ref{sec:turb_acc}).

\subsection{Turbulent acceleration, scattering and timescales} \label{sec:turb_acc}

In \cite{stackhouse2018spatially}, electrons are accelerated over an extended acceleration region using a spatially-dependent diffusion coefficient, where the spatial distribution is given as an exponential function. In order to explore different acceleration regions we adapt the acceleration diffusion coefficient in \cite{stackhouse2018spatially} to allow us to choose various spatial distributions. Thus, in this paper the acceleration diffusion coefficient $D(v,z)$ is given by:
\begin{equation}
    D(v,z) = \frac{v^2_{th}}{\uptau_{acc}}\left(\frac{v}{v_{th}}\right)^\alpha \times H(z)
   \label{Dvz}
\end{equation}
where $\uptau_{acc}$ is the chosen acceleration timescale, $\alpha$ is a constant controlling the velocity dependence, and $H(z)$ is a chosen spatial distribution of turbulence in the coronal apex and loop.

Similar to \cite{stackhouse2018spatially}, the acceleration timescale $\uptau_{acc}$ is considered as a multiple of the thermal collisional time, $\uptau_{c}$, given by:
\begin{equation}
    \uptau_{acc} = A_{acc} \uptau_c 
\end{equation}
where $A_{acc}$ is a constant set to either $A_{acc} =800, 1000, 2000, 4000$ or $6000$ in this paper. 

In our study, the spatial distribution $H(z)$ takes three different forms. Firstly, following \cite{stackhouse2018spatially}, we choose to model $H(z)$ as a Gaussian distribution:
\begin{equation}
    H(z) = \text{exp}\left(-\frac{z^2}{2\sigma ^2}\right) 
\end{equation}
centred at the coronal loop apex ($z=0\arcsec$), with the extent of the turbulent acceleration region controlled by $\sigma$, the standard deviation. Here, three different values of $\sigma$ are studied: $\sigma = 1\arcsec,3\arcsec,7\arcsec$ (referred to as $\sigma_1$, $\sigma_3$, $\sigma_7$, respectively). When modelled as a Gaussian, $H(z)$ will be referred to as $\rm{g}$ in the text. In all simulations, the loop length from the coronal apex to the corona-chromospheric boundary is set to $20\arcsec$ (with a total loop length of $40\arcsec$), thus the bulk of the acceleration region covers $\sigma/20\arcsec =5\%, 15\%, 35\%$ of the half-loop respectively\footnote{We note that the arcsecond is traditionally used to measure the source position and size from X-ray observations and values are calculated using a Sun-spacecraft distance of 1~AU but given values will vary for different Sun-spacecraft distances, e.g. SolO/STIX.}.

\citet{2021ApJ...923...40S} found a linear decrease in flare non-thermal broadening from the coronal looptop toward the footpoint/ribbon, particularly at times close to the peak in HXRs and thereafter. 
However, at early times in the same flare, the non-thermal velocity maps lacked a clear pattern and were quite chaotic. In order to account for these observations, two other spatial distributions: a linearly-decreasing distribution from the loop apex and a uniformly-random distribution, are studied here, given by:

\begin{equation}
H(z)=  \left (1 - \frac{|z|}{z_B} \right )
\label{Dvz_L}
\end{equation}
and 

\begin{equation}
H(z)= U[0,1]
\label{Dvz_R}
\end{equation}
where $z_B$ is the corona-chromospheric boundary set at $20\arcsec$ from the loop apex. These different spatial functions will be referred to as \textit{l} and \textit{r}, respectively. Both functions extend $3\arcsec$ from the apex, down the loop leg, after which point $H(|z|>3\arcsec) = 0$. When referring to the linear (\textit{l}) or random (\textit{r}) spatial functions, the value of $\sigma$ represents the extent of the acceleration region from the loop apex at $z = 0\arcsec$ along either sides of the loop.

We also test how the velocity dependency, controlled by the power index $\alpha$, changes the electron spectral and imaging properties. \cite{stackhouse2018spatially} use $\alpha = 3$ in order to compare to the leaky box model. To build upon this, we will be using three values of $\alpha= 2,3,4$ (referred to as $\alpha_{2}$, $\alpha_{3}$, and $\alpha_{4}$, respectively). When studying different $\alpha$, the spatial distribution will be modelled as $\rm{g}$ with $\sigma_{3}$. Similarly, when studying spatial extent and spatial function, $\alpha_{3}$ will always be used. 

Here, we scatter electrons (turbulent fluctuations) using an isotropic\footnote{The form of the underlying magnetic fluctuations is not well constrained in flare physics and hence, an isotropic scattering model is used.} pitch-angle diffusion coefficient \citep{1989ApJ...336..243S}, with the addition of our spatially-dependent term H(z):
\begin{equation}
D_{\mu\mu}(\mu,v,z)=v\frac{(1-\mu^{2})}{2\lambda_{ts}} \times H(z).
\end{equation}
Similar to \citet{2020A&A...642A..79J}, we define a scattering mean free path given by
\begin{equation}
\lambda_{ts}[E]=\lambda_{ts,0}\left(\frac{25}{E\rm{[keV]}}\right).
\label{tsmfp}
\end{equation}
Equation \ref{tsmfp} is taken from \cite{musset2018diffusive} and is determined empirically from X-ray imaging spectroscopy (mainly emitted by $<100$~keV electrons) and radio observations of gyrosynchrotron radiation (from $>100$~keV electrons). 

It is insightful to consider the different timescales involved in Equation \ref{time_independent_Fokker_planck}: collisional (thermalization) timescale $\uptau_{c}$, the collisional (pitch-angle scattering) timescale $\uptau_{c\mu} \approx \frac{2}{(1-\mu^2)}\frac{v^3}{\Gamma}$, the chosen acceleration timescale $\uptau_{acc}$, and the chosen turbulence scattering timescale $\uptau_{ts} = \lambda_{ts} / v$. Figure \ref{fig:1} shows the timescales involved for different electron energies.

For the turbulent scattering, we choose to model two cases: 
\begin{enumerate}
    \item[i)] Using Equation \ref{tsmfp} with $\lambda_{ts,0}=2\times10^{8}$ cm (following the observed result of \citealt{musset2018diffusive}). Turbulent scattering, $\uptau_{ts}$, dominates over collisional scattering, $\uptau_{\mu}$, for all energies $> 10$~keV (see Figure \ref{fig:1}c) with $\uptau_{ts}\sim0.02$~s for a 30~keV electron (or $\lambda_{ts}\sim2.3\arcsec$).
    \item[ii)] $\lambda_{ts}\rightarrow\infty$ i.e., creating a scenario where turbulent scattering is dominated by collisional scattering over all energies or occurs over approximately the same timescale.
\end{enumerate}  
Each case is chosen to cover a large range of possible underlying mechanisms of scattering and acceleration such that:
\begin{itemize}
\item In case i) $\uptau_{acc}>\uptau_{ts}$. Energy diffusion (acceleration) acts on a timescale greater than the scattering mechanism. This case will be referred to as `short timescale turbulent scattering'.
\item In case ii) $\uptau_{acc}$ is closer to $\uptau_{ts}=\uptau_{c\mu}$. Energy diffusion (acceleration) acts on a timescale approximately equal to the scattering mechanism. This case will be referred to as `without turbulent scattering'.
\end{itemize}

In Figure \ref{fig:1}c, using $\uptau_{acc} = [800, 1000,2000,4000,6000]\uptau_c$ we see that collisional timescales ($\uptau_{c}$ and $\uptau_{c\mu}$) dominate for low energy electrons ($<10$ keV). 

\subsection{Plasma and boundary conditions}
\label{plasma_conditions}

Here we accelerate electrons out of a thermal background already assumed to be heated to a high flare MK temperature. The source term $S_{F}(E,\mu,z)$ describes the input distributions for the energy, pitch angle and spatial position. Here, the electron distribution was input as an isothermal Maxwellian distribution with energies between $1-20$~keV. The injected pitch angle, $\mu$, has an isotropic distribution between $-1$ and $1$. The spatial distribution, $z$, was input as a Gaussian distribution centred at the loop apex $z = 0\arcsec$ with a standard deviation which equals $\sigma$ (or $3\arcsec$ for the $\textit{l}$ and $\textit{r}$ functions). 

The temperature and density of the flaring corona determine the collisional time. Here, the flaring coronal plasma is modelled as warm plasma in which the temperature and density remain uniform with values of $T = 20$~MK and $n = 3\times10^{10}$~cm$^{-3}$, with a Coulomb logarithm of $\text{ln}\Lambda = 20$. The chromospheric boundary is located $20\arcsec$ from the loop apex. At this boundary the temperature decreases to $T\approx0$ ($ 0.01$ K), ensuring a cold target, and the density increases to $n = 10^{12}$~cm$^{-3}$. It then exponentially-increases to photospheric densities ($\approx10^{17}$~cm$^{-3}$) over a scale height of $130$~km (as in \citealt{2012ApJ...752....4B,2019ApJ...880..136J}). Here, the Coulomb logarithm is reduced to $\text{ln}\Lambda = 7$. 

Acceleration can only occur in the corona and the simulation ends when all electrons leave the warm coronal plasma and enter the cold chromosphere. Here, they continuously lose energy, fall below $< 1$ keV and are removed from the simulation. In the corona, small acceleration timescales may result in the electrons continuously gaining energy. We only study electrons between $1$ and $100$~keV here, which is suitable for most flare X-ray observations (e.g., STIX only observes X-rays up to $\sim100$~keV). Thus, an upper energy boundary removes electrons with energies greater $1$ MeV, allowing the simulation to finish. This boundary is only applied when the electrons reach the chromosphere where electrons with energy $> 1$~MeV are considered lost in the Sun. An upper boundary of $1$~MeV allows the simulation to finish within a reasonable time whilst still removing high energy electrons. 

For comparison with observational data, we also consider the presence of a separate background thermal component in the coronal looptop,
\begin{equation}
    nVF_{th} = EM \sqrt{\frac{8}{\pi m_e}} \frac{E}{(k_BT)^{3/2}}e^{-E/k_BT}
    \label{eq:thermal_component}
\end{equation}
with $T=20$~MK as defined above. From comparison with observations, we use various emission measure (EM) with $10^{49}$ cm$^{-3}$ as a maximum across a region of $-5\arcsec < z < 5\arcsec$. The addition of a looptop thermal component can change the range of values expected for each diagnostic, discussed in \S \ref{results}. 

\subsection{Simulation overview}

Overall, we will study the seven different acceleration regions listed in Table \ref{tab:my_label}. The left column states 
the acceleration region name,
the remaining 3 columns detail the properties of the region.
\begin{table}[]
    \centering
     \caption{The seven acceleration regions studied. When a Gaussian ($\rm{g}$) spatial function is used, the value of $\sigma$ represents the standard deviation of the Gaussian distribution. When a linear (\textit{l}) or random (\textit{r}) spatial function are used the value of $\sigma$ represents the extent of the acceleration region from the loop apex.}
    \begin{tabular}{c c c c}
    \hline
    \hline
     Acceleration  & Spatial & Spatial & Velocity \\
     Region Name& Extent & Function & Dependence \\
      \hline
      $\sigma_3,\rm{g},\alpha_3$  & $\sigma = 3\arcsec$ & $\text{Gaussian}$ & $\alpha = 3$ \\
      $\sigma_1,\rm{g},\alpha_3$  & $\sigma = 1\arcsec$ & $\text{Gaussian}$ & $\alpha = 3$ \\
      $\sigma_7,\rm{g},\alpha_3$ & $\sigma = 7\arcsec$ & $\text{Gaussian}$ & $\alpha = 3$ \\
      $\sigma_3,\textit{l},\alpha_3$ & $\sigma = 3\arcsec$ & $\text{Linear}$ & $\alpha = 3$ \\
      $\sigma_3,\textit{r},\alpha_3$  & $\sigma = 3\arcsec$ & $\text{Random}$ & $\alpha = 3$ \\
      $\sigma_3,\rm{g},\alpha_2$  & $\sigma = 3\arcsec$ & $\text{Gaussian}$ & $\alpha = 2$ \\
      $\sigma_3,\rm{g},\alpha_4$  & $\sigma = 3\arcsec$ & $\text{Gaussian}$ & $\alpha = 4$ \\
      \hline
    \end{tabular}
    \label{tab:my_label}
\end{table}
Firstly, our `control simulation' has a velocity dependency of $\alpha=3$, and a Gaussian spatial distribution with a spatial extent of $\sigma=3\arcsec$. This acceleration region will be referred to as $\sigma_{3},\rm{g},\alpha_{3}$ and all other simulation runs are compared against this control simulation i.e., if a simulation run is stated to have $\sigma_1$ then all other variable properties will remain default values, i.e., $\rm{g}$ and $\alpha_{3}$. The remaining 6 acceleration regions are outlined below: 
\begin{itemize}
    \item Two acceleration regions change the spatial extent to $\sigma = 1\arcsec$ or $\sigma = 7\arcsec$, known as $\sigma_1,\rm{g},\alpha_3$ and $\sigma_7,\rm{g},\alpha_3$, respectively. These two regions use a Gaussian spatial distribution and $\alpha = {3}$. 
    \item Two acceleration regions change the spatial function to a linear function \textit{l} or a random function \textit{r}, known as $\sigma_3,\textit{l},\alpha_3$ and $\sigma_3,\textit{r},\alpha_3$, respectively. These two regions use a spatial extent of $3\arcsec$ and $\alpha={3}$. 
    \item Two acceleration regions change the velocity dependence to $\alpha = 2$ or $\alpha = 4$, known as $\sigma_3,\rm{g},\alpha_{2}$ and $\sigma_3,\rm{g},\alpha_{4}$, respectively. These regions use a Gaussian spatial distribution with $\sigma = 3\arcsec$. 
\end{itemize}

The seven acceleration regions are studied for five acceleration timescales: $\uptau_{acc} = [800,1000,2000,4000,6000]\uptau_c$. We also run each simulation using shorter timescale scattering and without scattering (see case i and case ii in \S \ref{sec:turb_acc}), giving a total of 70 simulated regions.

\section{Results}\label{results}
In order to study how the properties of turbulent acceleration alter observable electron properties, we produce outputs for the density-weighted electron spectrum $nVF$ [electrons $\text{ s}^{-1} \text{ cm}^{-1} \text{ keV}^{-1}$] \cite[e.g.,][]{2003ApJ...595L.115B}, a quantity that is directly linked to X-ray observations without making assumptions about the transport processes occurring in the corona. For a simplistic angle-integrated case, the relation between the X-ray distribution $I(\epsilon)$ and $nVF$ is given by
\begin{equation}
I(\epsilon)=\frac{1}{4\pi R^{2}}\int_{E=\epsilon}^{\infty}nVF(E)\,Q(E,\epsilon)dE
\end{equation}
where $\epsilon$ is the photon energy, $R$ is the Sun-Earth (or Sun-observation) distance and $Q$ is the angle-integrated bremsstrahlung cross section \citep{1959RvMP...31..920K}.

Most solar flare spectra are steeply decreasing power laws and the spectral index of the X-ray distribution is one parameter used to diagnose the properties of the accelerated electron population. Here, we do not inject an accelerated electron population but electrons are accelerated directly out of a thermal population. However, in a standard thick target model, an injected spectral index $\delta$ is often discussed. Instead, we use the spectral index $\delta_{nVF}$ of the emitting electron spectrum $nVF$. In a simple thick-target model, we would expect the inferred spectral index of an injected electron distribution $\delta \approx \delta_{nVF} + 2$ and the X-ray photon index $\gamma \approx \delta_{nVF} +1$. However, it is well-known that transport effects can increase or decrease this difference. For example, a small spectral difference between the X-ray coronal looptop (LT) source and the X-ray footpoint (FP) source ($\delta^{LT}_{nVF}-\delta^{FP}_{nVF}<2$) is consistent with the confinement of electrons in the corona \citep[e.g.,][]{2012ApJ...748...33C}.

\begin{figure}
    \centering
    \includegraphics{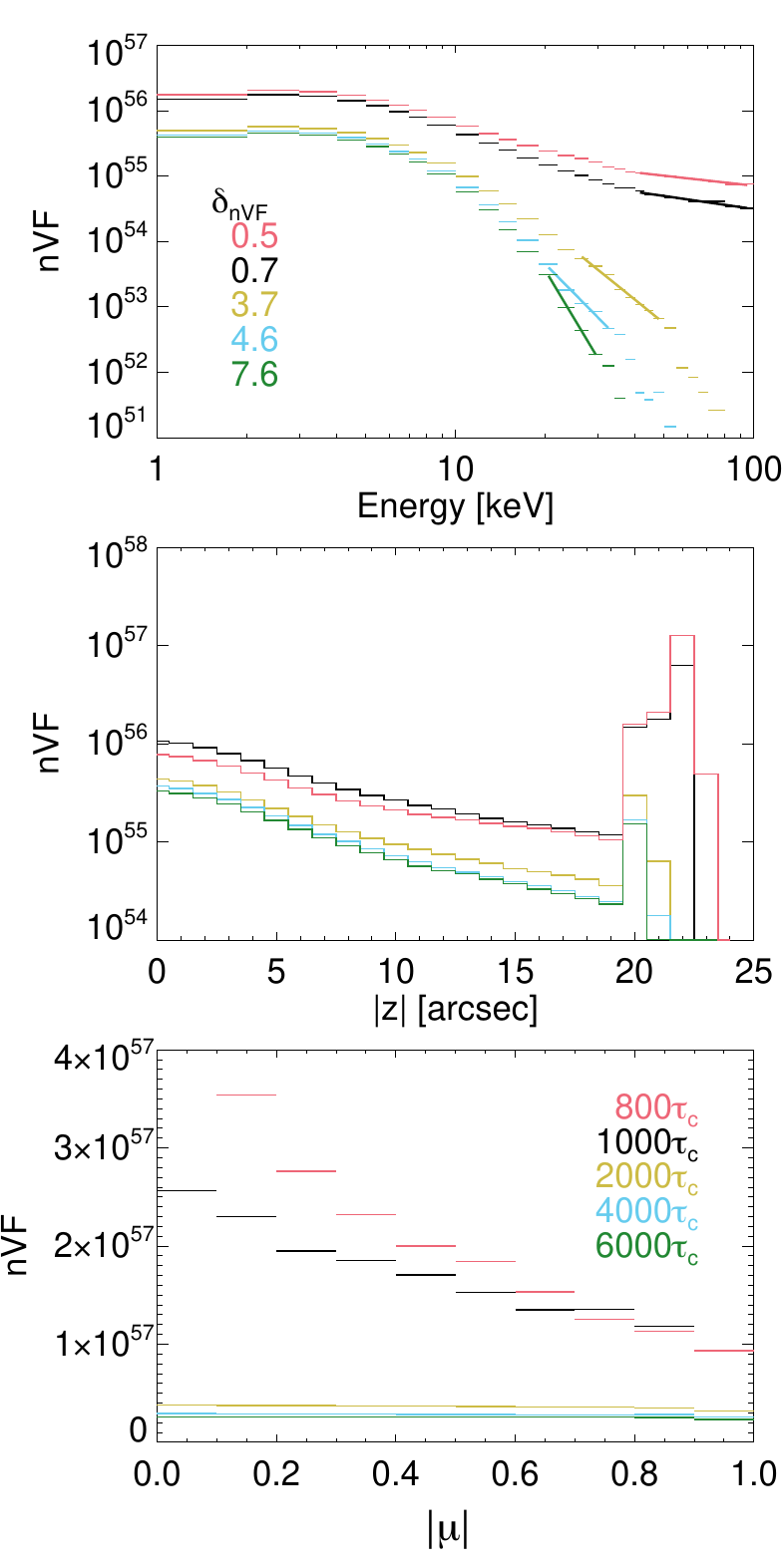}
    \caption{Example $nVF$ [units: electrons s$^{-1}$ cm$^{-2}$ keV$^{-1}$] distributions versus energy $E$ (top), space $|z|$ (middle) and pitch-angle $|\mu|$ (bottom). This example is the control simulation $\sigma_3,\rm{g},\alpha_3$, for different acceleration times $\uptau_{acc}$. Variations in acceleration properties (space, time) lead to distinctive changes (diagnostics) in all observables (spectra), imaging (space) and directivity (pitch-angle).}
    \label{fig:2}
\end{figure}

Figure \ref{fig:2} shows one example simulation output; $nVF(E,\mu,z)$ versus $E$, $|\mu|$, and $|z|$. This simulation output is created using our control simulation, $\sigma_3,\rm{g},\alpha_3$, for acceleration times of $\uptau_{\rm acc}=[800,1000,2000,4000,6000]\uptau_{c}$. 

The top panel of Figure \ref{fig:2} depicts the full (space and angle integrated) flare spectra $nVF$ versus $E$. A thermal part is dominant at energies below $\approx20$~keV. This thermal component is related to the studied electron population only (e.g., thermalization). Where possible, the addition of a separate background thermal component, $nVF_{th}$ in Equation \ref{eq:thermal_component}, will be examined in regards to how it will affect the described diagnostic tools.

The second row in Figure \ref{fig:2} plots $nVF$ versus $|z|$ (energy and angle integrated) for the different $\uptau_{acc}$. 
For all acceleration timescales shown in Figure \ref{fig:2} there is a peak in $nVF$ at the coronal looptop, centred at the apex $z = 0\arcsec$. A secondary peak in $nVF$ occurs in the chromosphere. 
The position of the peak in the chromosphere changes with $\uptau_{acc}$ (this will be discussed further in \S \ref{sect:footpoints}). Furthermore, as $\uptau_{acc}$ decreases, $nVF$ in the chromosphere increases, as expected.

The third row in Figure \ref{fig:2} plots $nVF$ versus $|\mu|$ (energy and space integrated) for the different $\uptau_{acc}$.  The initial electron $\mu$ distribution is isotropic and in the absence of anisotropic scattering mechanisms, we would expect the electron distribution to remain isotropic, which is true for the majority of simulated cases. However, in low $\uptau_{acc}$ cases ($800\uptau_{c}$ (pink) and $1000\uptau_{c}$ (black)), the distribution can move away from isotropic\footnote{In this particular case, without short timescale turbulent scattering, the lack of collisional scattering at high energies and the rapid acceleration in these cases leads to electrons becoming trapped at certain angles. Electrons trapped at small angles stream to the chromosphere while electrons trapped at angles closer to $90^{\circ}$ spend longer in the corona.} and peak closer to $90^{\circ}$. 
We do not discuss the resulting directivity further in this study; this will form part of an ongoing study regarding the measurement of solar flare electron directivity from X-ray observations.

In the following subsections, we study spectral diagnostics; both spatially resolved and integrated, and spatial (imaging) diagnostics that can infer the properties of acceleration in an individual flare from X-ray observations alone. 

\subsection{Spectral and imaging diagnostics}
\label{sect:spectral}

\subsubsection{Full flare energy spectra}
The main diagnostic tool of past (e.g., RHESSI) and current (e.g., SolO/STIX) flare X-ray instrumentation is high resolution spectroscopy ($\sim$few keV, space-integrated and space-resolved) constraining the bulk of the accelerated electron properties to-date. Similar to \citet{2018A&A...612A..64S}, the top panel of Figure \ref{fig:2} shows how the flare spectra change due to varying $\uptau_{acc}$ with the shape of the spectrum changing significantly from $\uptau_{acc} = 800\uptau_c$ to $\uptau_{acc} = 6000\uptau_c$. When the acceleration timescale is high, i.e $\uptau_{acc} = 6000\uptau_c$, $nVF$ remains mostly thermal. However, as $\uptau_{acc}$ decreases, $nVF$ increases and a non-thermal tail forms (harder spectrum). As expected, 
the spectral index $\delta_{nVF}$ increases as acceleration timescale increases ranging from $\delta_{nVF}=0.5$ at $\uptau_{acc} = 800\uptau_c$ to $\delta_{nVF}=7.6$ at $\uptau_{acc} = 6000\uptau_c$\footnote{We note that in a traditional thick-target analysis this leads to an injected $\delta\approx0.5+2=2.5$ and $\delta\approx7.6+2=9.6$ respectively.}. 
In Figure \ref{fig:2} (top), a single power law is fitted to the data over an appropriate energy range to find the spectral index (shown by the solid lines).

\begin{figure*}
    \centering
    \includegraphics[scale = 0.43]{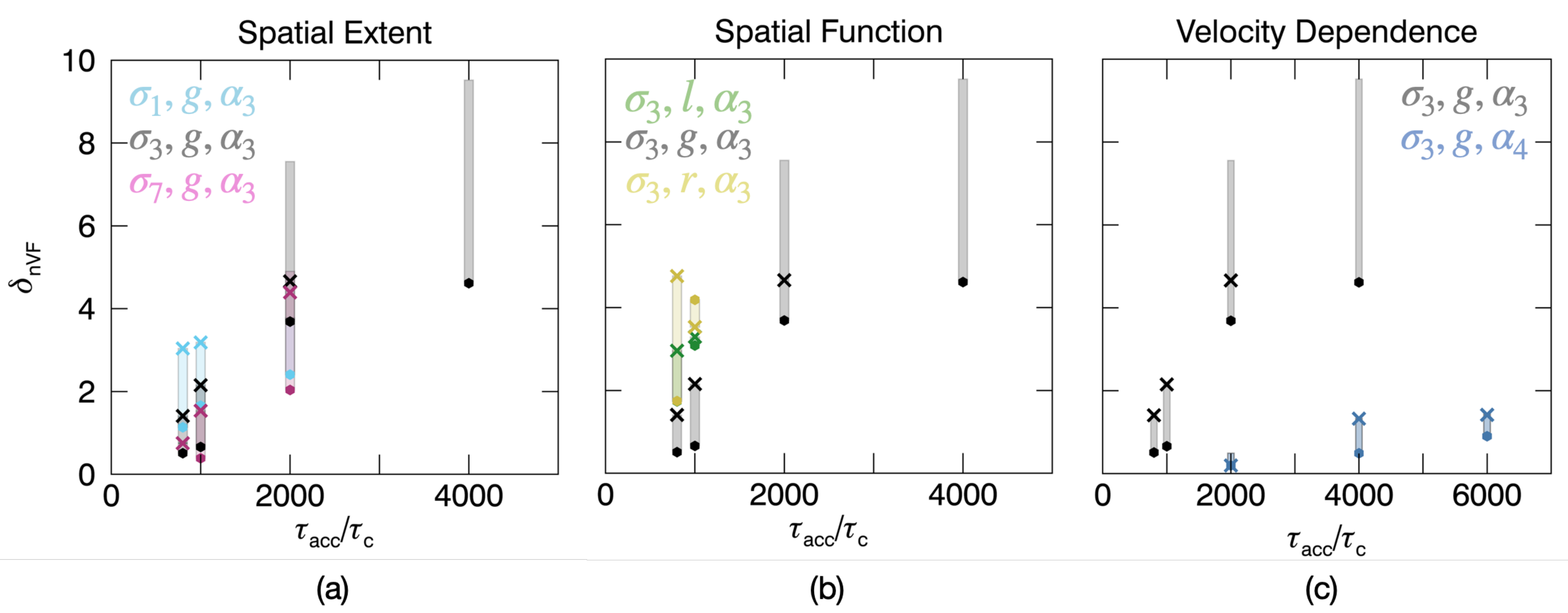}
    \caption{Full flare spectral index $\delta_{nVF}$ versus $\uptau_{acc}$ for different: (a) spatial extents ($\sigma_1,\rm{g},\alpha_3$ (light blue), $\sigma_7,\rm{g},\alpha_3$ (pink)); (b) spatial functions ($\sigma_3,\textit{l},\alpha_3$ (green), $\sigma_3,\textit{r},\alpha_3$  in green (yellow)); (c) velocity dependence ($\sigma_3,\rm{g},\alpha_4$ (dark blue)). Our control simulation $\sigma_3,\rm{g},\alpha_3$ is shown in each graph (black/gray). Simulations with and without the inclusion of shorter timescale turbulent scattering are shown with crosses and circles. Each shaded region indicates the possible range of $\delta_{nVF}$ when considering the addition of a background thermal component (up to a maximum value of EM$=10^{49}$~cm$^{-3}$).}
    \label{fig:3}
\end{figure*}

Figure \ref{fig:3} shows the full-flare $\delta_{nVF}$ versus $\uptau_{acc}$, for all spatial extents (\ref{fig:3}a), spatial functions (\ref{fig:3}b), and velocity dependencies (\ref{fig:3}c). Simulations `with' and `without' short timescale turbulent scattering, case i) and ii), are shown with crosses and circles respectively. Shaded rectangles indicate the addition of a background thermal component, $nVF_{th}$ (with a maximum ${\rm EM}=10^{49}$~cm$^{-3}$), that may change the inferred value of $\delta_{nVF}$.
For each spectra, the spectral index $\delta_{nVF}$ is determined from $E = 20$~keV or greater (to avoid the influence of the thermal part at lower energies) until the highest suitable energy, with a maximum value of $E = 100$~keV. We choose this maximum value for two reasons: (1) most simulation outputs are noisy after this energy and (2) most solar flare spectra (e.g. RHESSI, STIX) are fitted between this range above the background. If the highest suitable energy was $< 30$ keV $\delta_{nVF}$ was fitted from a minimum of $15$~keV and the fits were studied carefully by eye. Several simulations have an increasing energy spectrum as a result of very high acceleration (i.e $\sigma_3,\rm{g},\alpha_{4} $ at $\uptau_{acc} = 800\uptau_c$), and the spectral index of these simulations has been removed from Figure \ref{fig:3} since we do not see increasing spectra in solar flare data. Similarly, several simulations have only thermal spectra, where $\delta_{nVF}$ could not be determined. 

Considering Figure \ref{fig:3}, as $\delta_{nVF}$ increases, $\uptau_{acc}$ increases for all simulations. For the simulations studied, $\sigma_3,\rm{g},\alpha_{4}$ provides a lower boundary on $\delta_{nVF}$ at a given $\uptau_{acc}$, producing the hardest spectra (Figure \ref{fig:3}c). Increasing the velocity dependency increases the spectral index such that $\delta_{nVF}[\alpha_3 > \alpha_4]$ for a given acceleration timescale.  No values for $\sigma_3,\rm{g},\alpha_{2}$ are included, as the energy spectra was thermal for all acceleration timescales. 

Figure \ref{fig:3}a shows that as the spatial extent increases, $\delta_{nVF}$ decreases, such that $\delta_{nVF}$[$\sigma_1 > \sigma_3 > \sigma_7$], as expected, since electrons experience more acceleration in an acceleration region of greater spatial extent, leading to harder spectra. 
In contrast, changing $H(z)$ (Figure \ref{fig:3}b) does not noticeably change $\delta_{nVF}$ with $\uptau_{acc}$. In general, both $\textit{l}$ and $\textit{r}$ with extent $3\arcsec$ ($\sigma_3,\textit{l},\alpha_3$ and $\sigma_3,\textit{r},\alpha_3$, respectively) produce larger $\delta_{nVF}$ than $\rm{g}$ of the same spatial extent ($\sigma_3,\rm{g},\alpha_3$) and a more apt comparison may be with a $\rm{g}$ with a smaller spatial extent, i.e $\sigma_1,\rm{g},\alpha_3$.

The addition of short timescale turbulent scattering and/or a background thermal component increases the spectral index. Generally, as the acceleration timescale increases, the range of the spectral index increases. Whilst it is still possible to distinguish between different velocity dependencies at larger $\uptau_{acc}$, it is difficult to distinguish between other parameters such as spatial extent.  


\subsubsection{Looptop and footpoint energy spectra}
\begin{figure*}[t!]
    \centering
    \includegraphics[scale = 0.85]{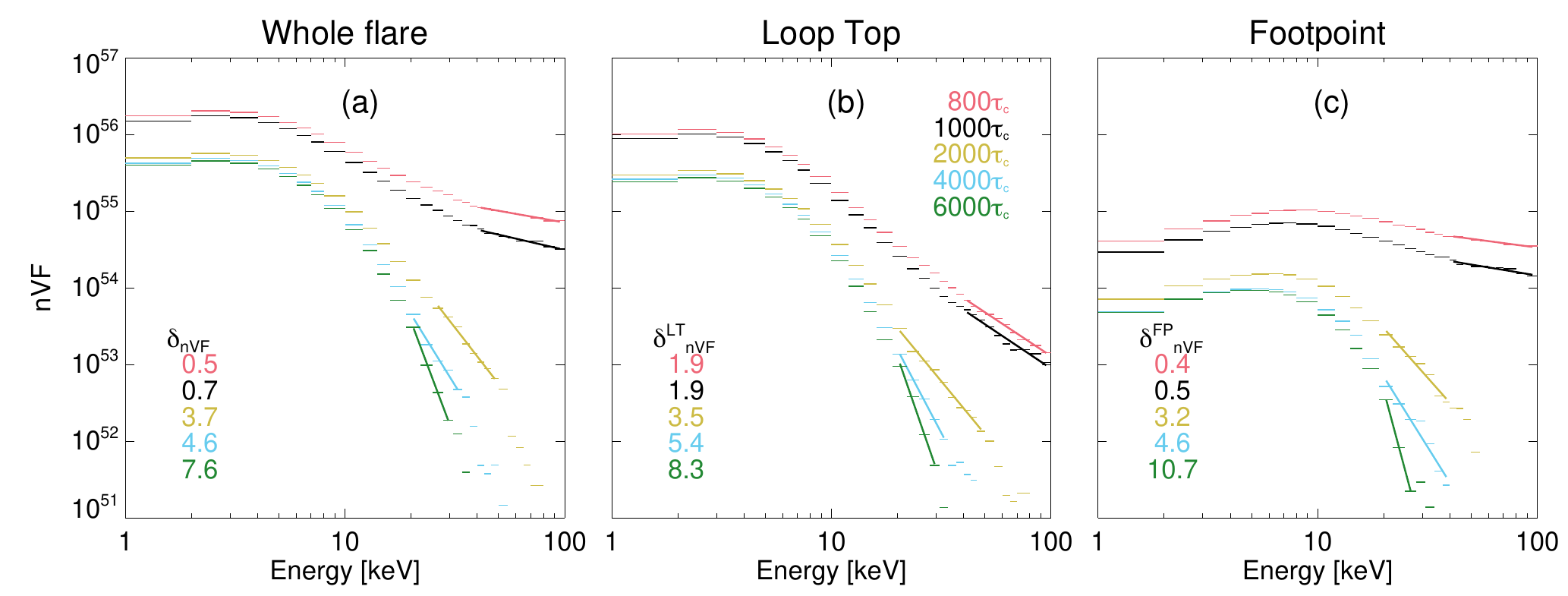}
    \includegraphics[scale = 0.41]{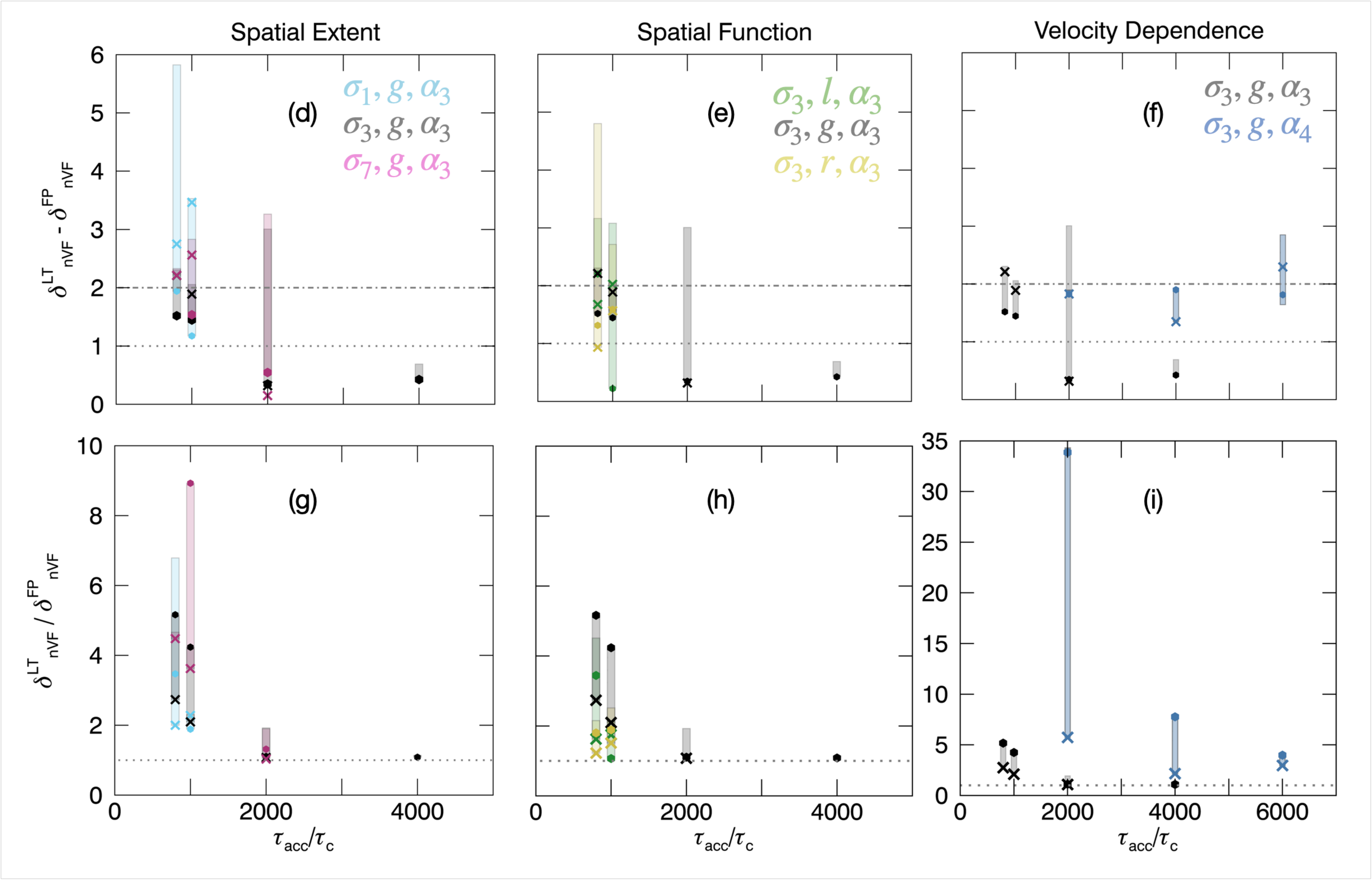}
    \caption{Panels a-c: Control simulation energy spectra ($nVF$ [units: electrons s$^{-1}$ cm$^{-2}$ keV$^{-1}$] versus $E$) for: (a) the full flare, (b) coronal looptop (LT) only, and (c) footpoint (FP) only (no background thermal component), all values of $\uptau_{acc}$ are shown in different colors given in the legend. 
    (Panels d-f) Spectral index difference $\delta^{LT}_{nVF} - \delta^{FP}_{nVF}$ and (panels g-i) Spectral index ratio $\delta^{LT}_{nVF}/\delta^{FP}_{nVF}$ vs. $\uptau_{acc}$ for simulations changing: (d and g) spatial extent, (e and h) spatial function, and (f and i) velocity dependence. Simulations with and without the inclusion of shorter timescale turbulent scattering are shown with crosses and circles. Each shaded region indicates the possible range of $\delta_{nVF}$ when considering the addition of a background thermal component (up to a maximum value of EM$=10^{49}$~cm$^{-3}$).}
\label{fig:4}
\end{figure*}

Imaging spectroscopy (spatially resolved spectra) is an essential tool for constraining the properties of electron acceleration and transport in flares. In Figures \ref{fig:4}a-c we plot examples of spatially resolved energy spectra showing separate spectra for the coronal looptop (\ref{fig:4}b) and chromospheric footpoints (\ref{fig:4}c) alongside the full flare integrated spectrum (\ref{fig:4}a). A comparison of the spectral indices for the full flare $\delta_{nVF}$, looptop $\delta_{nVF}^{LT}$ and footpoint $\delta_{nVF}^{FP}$ sources are shown in the legends. Similar to observational imaging spectroscopy, we define the looptop region at $-5\arcsec<z<5\arcsec$, and the footpoints sources at $ 20\arcsec< |z| < 30\arcsec$ \,\footnote{Here, the shown footpoint spectra is the mean spectra of both footpoint spectra.}.
As in Figure \ref{fig:2}, the example spectra in Figure \ref{fig:4}a-\ref{fig:4}c shows the control simulation for different $\uptau_{acc}$. A power law fit (thick solid line) is used to find the spectral index.

In general, for each simulation $\delta_{nVF}$ is harder than the looptop spectral index $\delta_{nVF}^{LT}$. Similarly, the footpoint spectral index $\delta_{nVF}^{FP}$ is harder than $\delta_{nVF}$, such that, $\delta^{FP}_{nVF} > \delta_{nVF} > \delta^{LT}_{nVF}$, as expected. For example, the control simulation with $\uptau_{\rm acc}=1000\uptau_c$ gives $\delta^{FP}_{nVF}=0.5 > \delta_{nVF}=0.7 > \delta^{LT}_{nVF}=1.9$.


Figure \ref{fig:4}d, \ref{fig:4}e, and \ref{fig:4}f plot the difference $\delta^{LT}_{nVF} - \delta^{FP}_{nVF}$ for different spatial extents (\ref{fig:4}d), spatial functions (\ref{fig:4}e), and velocity dependencies (\ref{fig:4}f). 
The larger the value of $\alpha$ (velocity dependence), the greater the spectral difference $\delta^{LT}_{nVF} - \delta^{FP}_{nVF}$ with $\sigma_3,\rm{g},\alpha_{4}$ showing $\delta^{LT}_{nVF} - \delta^{FP}_{nVF}\approx2$ for all $\uptau_{acc}$. In this case, electrons are accelerated rapidly and behave similar to an injected distribution in the corona that streams to the chromosphere, approximately following the simple thick-target expectation.
The smaller the spatial extent, the larger the spectral difference $\delta^{LT}_{nVF} - \delta^{FP}_{nVF}$ since electrons escape the acceleration region faster. This may also explain why the $\textit{l}$ and $\textit{r}$ spatial functions generally have larger spectral differences compared to $\rm{g}$, since the spatial extent is $3\arcsec$ (see Figure \ref{fig:1}b). 

Previous work \citep[e.g,][]{2012ApJ...748...33C} shows that a spectral index difference less than 2 is often consistent with the confinement of electrons in the corona. Firstly, this is mirrored in our results when the acceleration becomes less efficient. Also, we do not need to necessarily invoke different turbulent spectra or exotic plasma waves that preferentially accelerate electrons at $90^{\circ}$ \citep{2012ApJ...748...33C}, for such confinement. For example, as $\uptau_{acc}$ increases, the spectral index difference is smaller for all simulation runs apart from the extreme $\sigma_3,\rm{g},\alpha_4$ case that has efficient acceleration across all $\uptau_{acc}$. Our different scattering cases i) and ii) also change the spectral index difference. When turbulent scattering acts on longer timescales (circles in Figure \ref{fig:4}) closer to the collisional time, the spectral difference is $\lesssim2$ for all cases. When the turbulent scattering acts on shorter timescales (crosses), the spectral index difference tends to increase slightly at low $\uptau_{acc}$, which is mainly due to the formation of steeper spectra in these cases. 
Further, the spectral index difference increases when a larger background thermal component is present since the non-thermal component in the corona is hidden by the large (steep) thermal background. 

We also study the usefulness of the ratio diagnostic $\delta^{LT}_{nVF}/\delta^{FP}_{nVF}$ with $\uptau_{acc}$, seen in Figures \ref{fig:4}g, \ref{fig:4}h, and \ref{fig:4}i. 
For all acceleration regions,  $\delta^{LT}_{nVF}/\delta^{FP}_{nVF}$ decreases as $\uptau_{acc}$ increases. When $\uptau_{acc} \leq 1000\uptau_c$ the ratio $\delta^{LT}_{nVF}/\delta^{FP}_{nVF} > 1$. At high acceleration timescales ($\uptau_{acc} \geq 2000\uptau_c$), the spectrum is mainly thermal, and the ratio begins to flatten remaining at $\sim 1$.

Shorter timescale turbulent scattering generally leads to smaller spectral index ratios (again due to the steeper footpoint spectra), as does the addition of a background thermal component with increasing EM. Furthermore, the greater the velocity dependence or spatial extent, the greater the ratio. Once again, the spatial functions are harder to separate. However, $\rm{g}$ often has a greater ratio than \textit{l} and \textit{r}. Thus in general, the more efficient the acceleration in the loop, the larger the ratio $\delta^{LT}_{nVF}/\delta^{FP}_{nVF}$. 

Although differences in spectral index between looptop and footpoint sources are often discussed in observational work \citep[e.g., ][]{2006A&A...456..751B,2012ApJ...748...33C,2013A&A...551A.135S}, here we find that the ratio of spectral indices $\delta^{LT}_{nVF}/ \delta^{FP}_{nVF}$ is the more reliable diagnostic for indicating the acceleration timescale when there is a small background thermal component. Whereas, the spectral difference is a stronger diagnostic for the velocity dependence. 


\subsection{Energy-averaged flux ratios}
\begin{figure*}
    \centering
    \includegraphics[scale = 0.85]{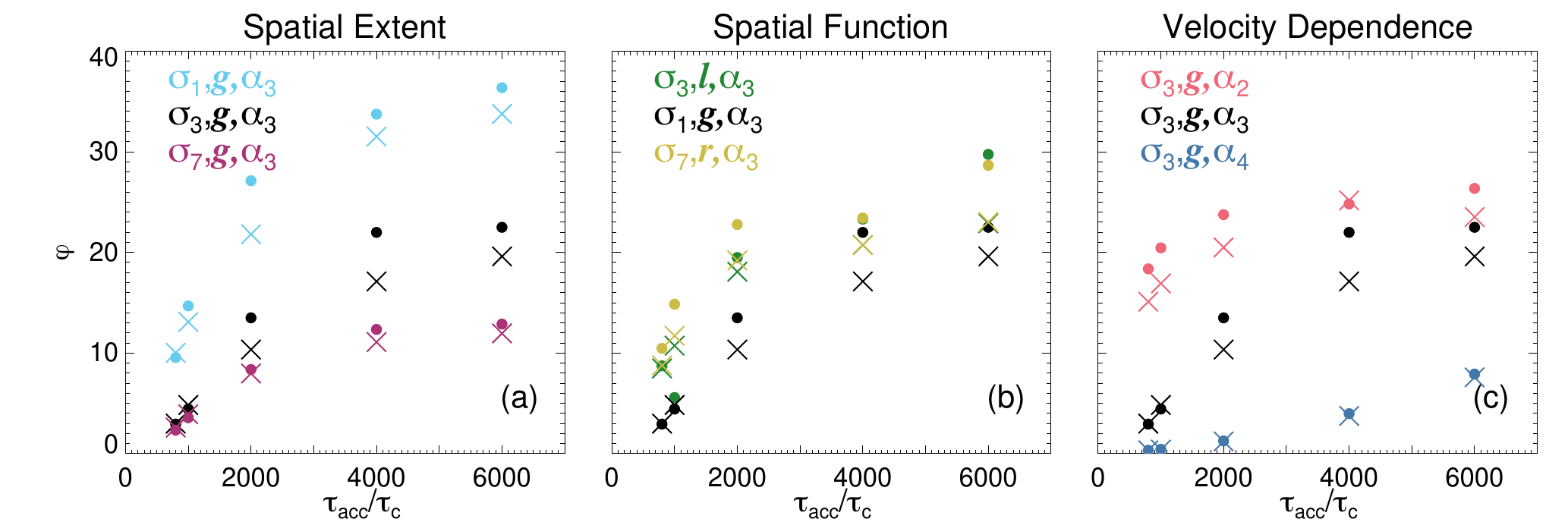}
    \includegraphics[scale = 0.85]{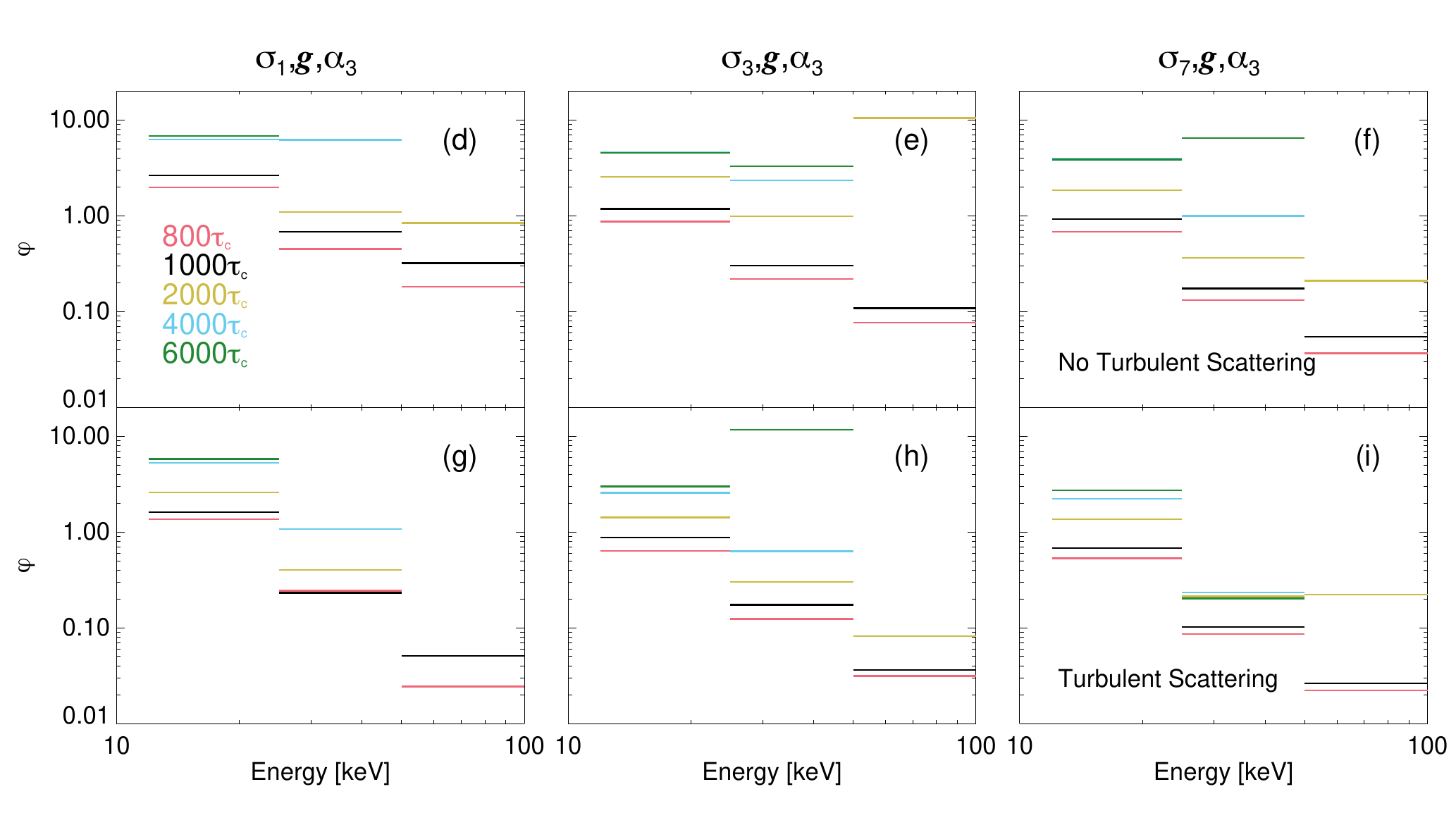}
    \caption{(Panels a-c) ratio of looptop $nVF$ and footpoint $nVF$, defined as $\varphi= nVF(-5 \arcsec<z <5\arcsec)/nVF(20 \arcsec < |z|<  30\arcsec)$ (energy-integrated), versus $\uptau_{acc}$ for simulations which change: (a) the spatial extent, (b) spatial function, and (c) velocity dependence of the acceleration region. Simulations with and without the inclusion of shorter timescale turbulent scattering are shown with crosses and circles. 
    (Panels d-i) show an example of energy-dependent $\varphi$ versus energy for different spatial extents, i.e, (d and g) $\sigma_1,\rm{g},\alpha_3$, (e and h) $\sigma_3,\rm{g},\alpha_3$, and (f and i) $\sigma_7,\rm{g},\alpha_3$. These results do not include an additional background thermal component.}
    \label{fig:5}
\end{figure*}

Figures \ref{fig:5}a, \ref{fig:5}b, and \ref{fig:5}c plot the ratio of looptop $nVF$ to footpoint $nVF$ over all energies defined as,
\begin{equation}
    \varphi = \frac{nVF( - 5\arcsec < z <5\arcsec)}{nVF( 20 \arcsec< |z| <30\arcsec )} \, ,
\end{equation}
against $\uptau_{acc}$ for different spatial extents (\ref{fig:5}a), spatial functions (\ref{fig:5}b), and velocity dependencies (\ref{fig:5}c). Simulations with and `without' shorter timescale turbulent scattering are shown with crosses and circles, respectively. We exclude a background thermal component from Figure \ref{fig:5} as this component can dominate this particular diagnostic by several orders of magnitude for large EM. Thus, this diagnostic will not be suitable for observations which have a large background thermal component. If $\varphi>1$, looptop emission dominates and if $\varphi<1$, footpoint emission dominates. For each acceleration region studied, $\varphi$ increases as $\uptau_{acc}$ increases, with $\varphi>1$ for all $\uptau_{acc}$, except $\sigma_3,\rm{g},\alpha_{4}$ at $\uptau_{acc} \leq 1000\uptau_c$. 
Thus, for almost all simulations, the looptop emission (consisting of lower energy emission) dominates regardless of the acceleration timescale, i.e., due to decreasing energy spectra in solar flares. 

For suitable flares with small background EM (possibly so-called `cold' flares e.g., \citealt[][]{2020ApJ...890...75M} or microflares e.g., \citealt[][]{2020ApJ...891L..34G}), the spatial extents and velocity dependencies of the acceleration region are clearly separated by $\varphi$, such that $\varphi[\alpha_2 > \alpha_3 > \alpha_4]$ and  $\varphi[\sigma_1 > \sigma_3 > \sigma_7]$. Moreover, there is even a slight separation between different spatial functions, with $\varphi[ \textit{r} > \textit{l} > \rm{g}]$. The difference between $\textit{r}$ and $\textit{l}$ may be due to the inhomogeneous acceleration when the spatial function is random, $\textit{r}$. As a result there are locations in the loop where electrons experience no acceleration. In comparison, electrons moving within $\textit{l}$ experience (decreasing) acceleration at all locations within the region extent. 

Simulations `without' turbulent scattering can lead to smaller $\varphi$ since electrons `locked' at $\approx90^{\circ}$ can experience greater trapping (and produce more coronal emission) even compared to the confinement produced by short-timescale scattering, case i). Here, this results in simulations without shorter timescale turbulent scattering, case ii), having a value of $\varphi$ that is approximately 1.0 - 1.5 times greater. 

\subsubsection{Energy-dependent flux ratios}
Instruments such as RHESSI and STIX provide us with spatially resolved data of solar flares in several energy bins.
Figures \ref{fig:5}d-\ref{fig:5}i  plot energy-dependent $\varphi$ versus $E$. For realistic comparison with imaging data, we use the following energy bins: $E = 3-6~\text{keV}, 6-12~\text{keV},12-25~\text{keV}, 25-50~\text{keV}, 50-100~\text{keV}$. Figure \ref{fig:5} shows simulations without turbulent scattering (\ref{fig:5}d-\ref{fig:5}f) and with short-timescale  turbulent scattering (\ref{fig:5}g-\ref{fig:5}i).

As energy increases, $\varphi$ decreases. Thus, as $E$ increases the footpoint emission becomes more dominant, as expected. The examples shown in Figures \ref{fig:5}d-\ref{fig:5}i compare different spatial extents,  $\sigma_{1},\rm{g},\alpha_3$ (\ref{fig:5}d and \ref{fig:5}g), $\sigma_{3},\rm{g},\alpha_3$ (\ref{fig:5}e and \ref{fig:5}h), and $\sigma_{7},\rm{g},\alpha_3$ (\ref{fig:5}f and \ref{fig:5}i) for all $\uptau_{acc}$\footnote{At higher energies, data may be missing at larger acceleration timescales.}. We choose to compare spatial extent $\sigma$, since this property produces a larger variation in $\varphi$ with $\uptau_{acc}$ as shown in Figures \ref{fig:5}a - \ref{fig:5}c. For a given $\sigma$, we see differences in $\varphi$ with energy for different $\uptau_{acc}$. For example, at lower $\uptau_{acc}$ and for a smaller spatial extent, we see larger $\varphi$, as electrons experience less acceleration. 
For the other studied variables (not shown), such as velocity dependence, as $\alpha$ increases we see a larger spread in $\varphi$ with $\uptau_{acc}$ for a given energy. Acceleration regions with $\textit{r}$ show slightly less variation in $\varphi$ with $\tau_{acc}$, similar to simulations with $\alpha_2$. Whereas, the $\textit{l}$ and $\rm{g}$ distributions show a larger range of $\varphi$ versus $\uptau_{acc}$ for each energy bin.

Shorter timescale turbulent scattering decreases energy dependent $\varphi$ by up to an order of magnitude, with this difference increasing with energy $E$, due to the decrease in higher energy footpoint emission. 

Adding a background thermal component increases $\varphi$ by several orders of magnitude (not shown). However, electrons with energy $\geq 25$~keV are not dominated by the thermal component at acceleration timescales $\leq 2000\uptau_c$. Overall, if a flare has a small background thermal component, $\varphi$ may help to constrain all properties of the acceleration region. However, if a flare has a large background thermal component, energy-independent $\varphi$ becomes obsolete and energy-dependent $\varphi$ is not useful for $E\leq 25$~keV.

\subsection{Changes in $nVF$ in space}

Figure \ref{fig:2} shows the energy- and angle-integrated spatial distribution for the control simulation for different $\uptau_{acc}$. At the coronal looptop, centred at $z = 0\arcsec$, $nVF$ generally increases as $\uptau_{acc}$ decreases. However, when acceleration is very large (small $\uptau_{acc}$), the coronal emission decreases with $\uptau_{acc}$. 
The chromospheric emission changes drastically as $\uptau_{acc}$ decreases, with $nVF$ increasing by up to two orders of magnitude, with greater $nVF$ at greater depths (more emission deeper in the chromosphere). 
Looking at velocity dependence, simulations which use $\alpha_{2}$ experience very little acceleration. For all $\uptau_{acc}$, the energy spectra are mainly thermal. This leads to spatial distributions with large $nVF$ at the loop apex compared to the footpoint (similar to $\sigma_3,\rm{g},\alpha_3$ at $\uptau_{acc} = 6000\uptau_c$ in Figure \ref{fig:2}). As $\uptau_{acc}$ decreases, $nVF$ in both the looptop and footpoints increases. 

Alternatively, for $\alpha_4$ and small $\uptau_{acc}$, the majority of the emission is concentrated to the chromospheric footpoints (similar to $\sigma_3,\rm{g},\alpha_3$ at $\uptau_{acc} = 800\uptau_c$ in Figure \ref{fig:2}). For $\alpha_4$, as $\uptau_{acc}$ decreases, $nVF$ in the looptop decreases for all values of $\uptau_{acc}$, which is in contrast to $\alpha_{2}$. Compared to $\alpha_2$, electrons move deeper into the chromosphere (greater $nVF$ at greater depths). This will be further discussed in \S \ref{sect:footpoints}.

Simulations with less overall acceleration, i.e., $\sigma_{1},\rm{g},\alpha_3$, produce spatial distributions similar to that of $\alpha_{2}$, for all $\uptau_{acc}$. Once again, simulations using $\textit{l}$ or $\textit{r}$ may be better compared to a $\rm{g}$ simulation using a smaller spatial extent, i.e $\sigma_1,\rm{g},\alpha_3$, than $\sigma_3,\rm{g},\alpha_3$. 

More interestingly, simulations using $\sigma_{3},\rm{g},\alpha_3$ and $\sigma_{7},\rm{g},\alpha_3$ show spatial distributions similar to simulations using either $\alpha_{2}$ and $\alpha_{4}$ depending on $\uptau_{acc}$. At large $\uptau_{acc} \geq 2000\uptau_c$, $\sigma_{3},\rm{g},\alpha_3$ and $\sigma_{7},\rm{g},\alpha_3$ have spatial distributions similar to $\alpha_{2}$, in which $nVF$ in the looptop increases as $\uptau_{acc}$ decreases and $nVF$ in the footpoints and looptops are of the same order of magnitude. 

Whereas, when $\uptau_{acc} < 1000\uptau_c$, $\sigma_{3},\rm{g},\alpha_{3}$ and $\sigma_{7},\rm{g},\alpha_{3}$ have spatial distributions similar to $\alpha_{4}$, such that $nVF$ in the looptop decreases as $\uptau_{acc}$ decreases and $nVF$ in the footpoints is up to two orders of magnitude greater than in the looptop.

The addition of the energy dependent shorter timescale turbulent scattering model reduces $nVF$ in both the corona and chromosphere. 
However, the width (i.e., depth of emission) in the chromosphere is not changed as significantly. 

\subsubsection{Energy-dependent spatial distribution}
\begin{figure*}[t!]
    \centering
    \includegraphics[scale = 0.85]{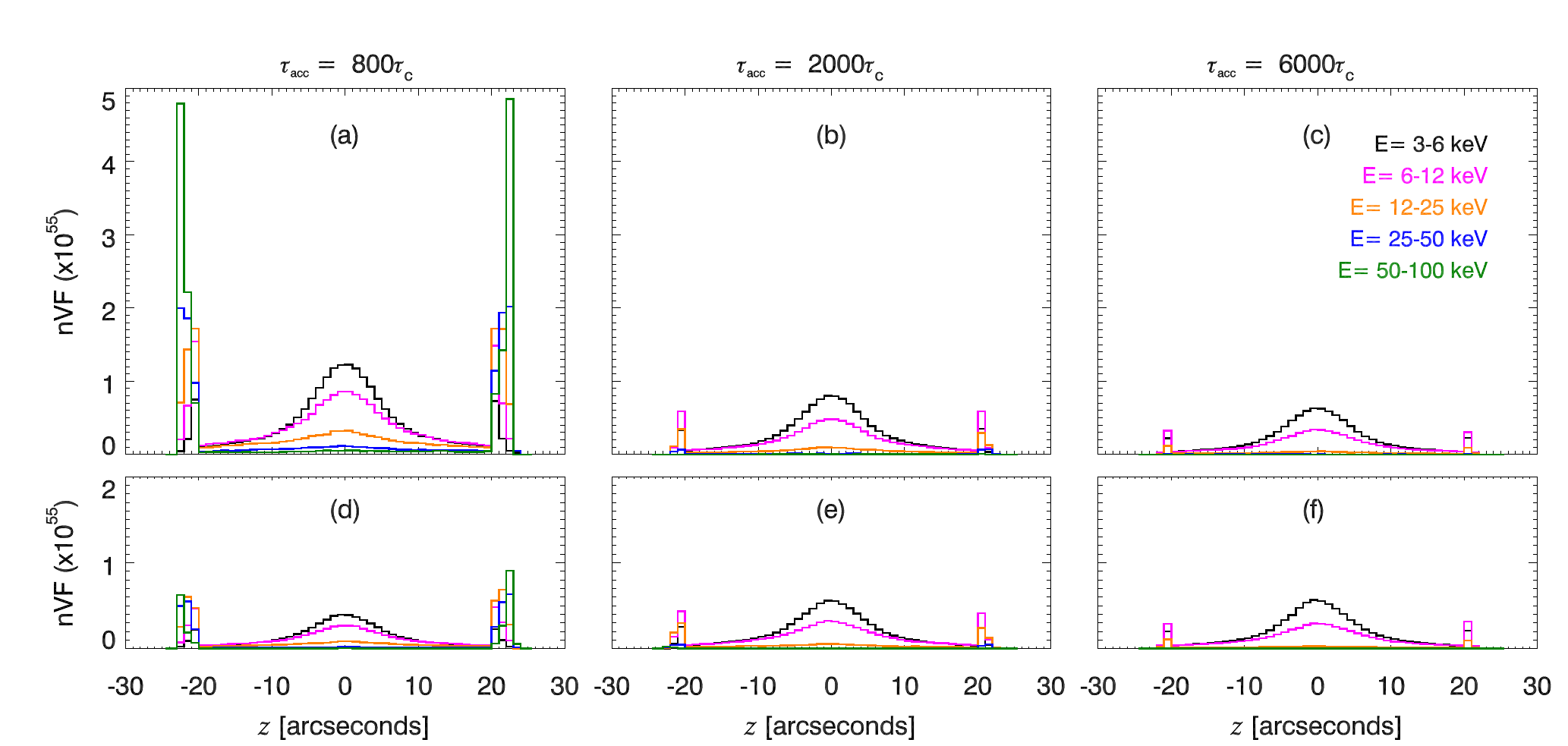}
    \includegraphics[scale = 0.55]{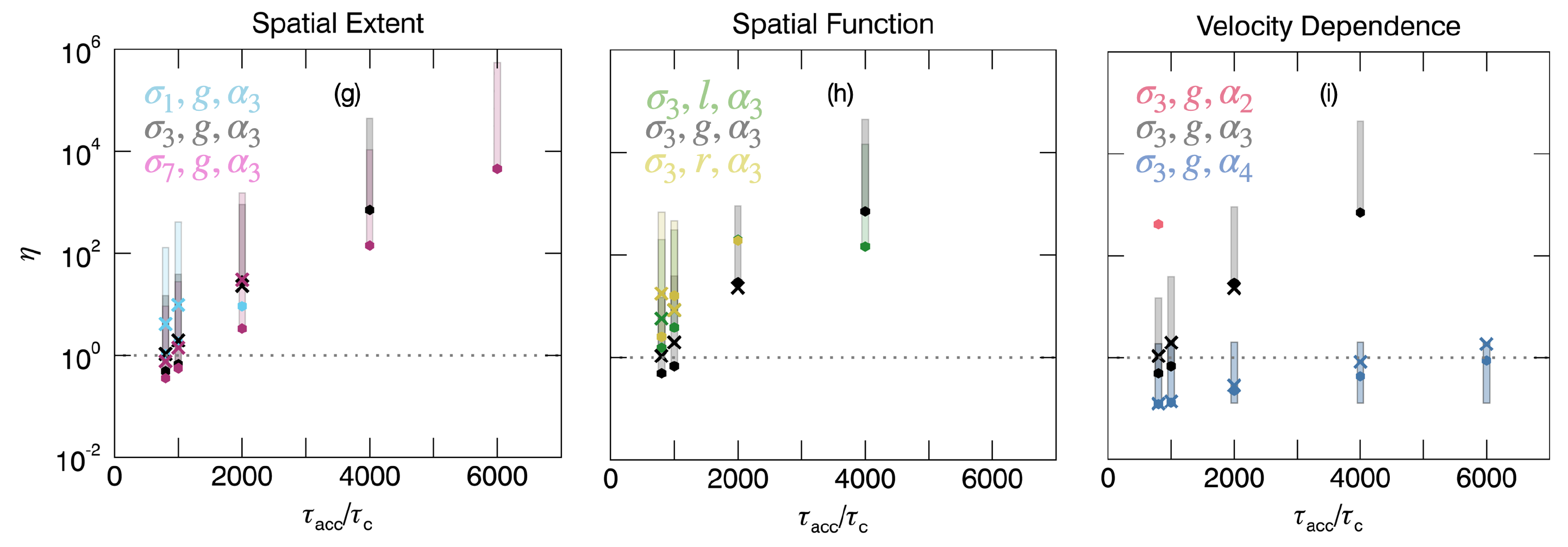}    
    \includegraphics[scale = 0.75]{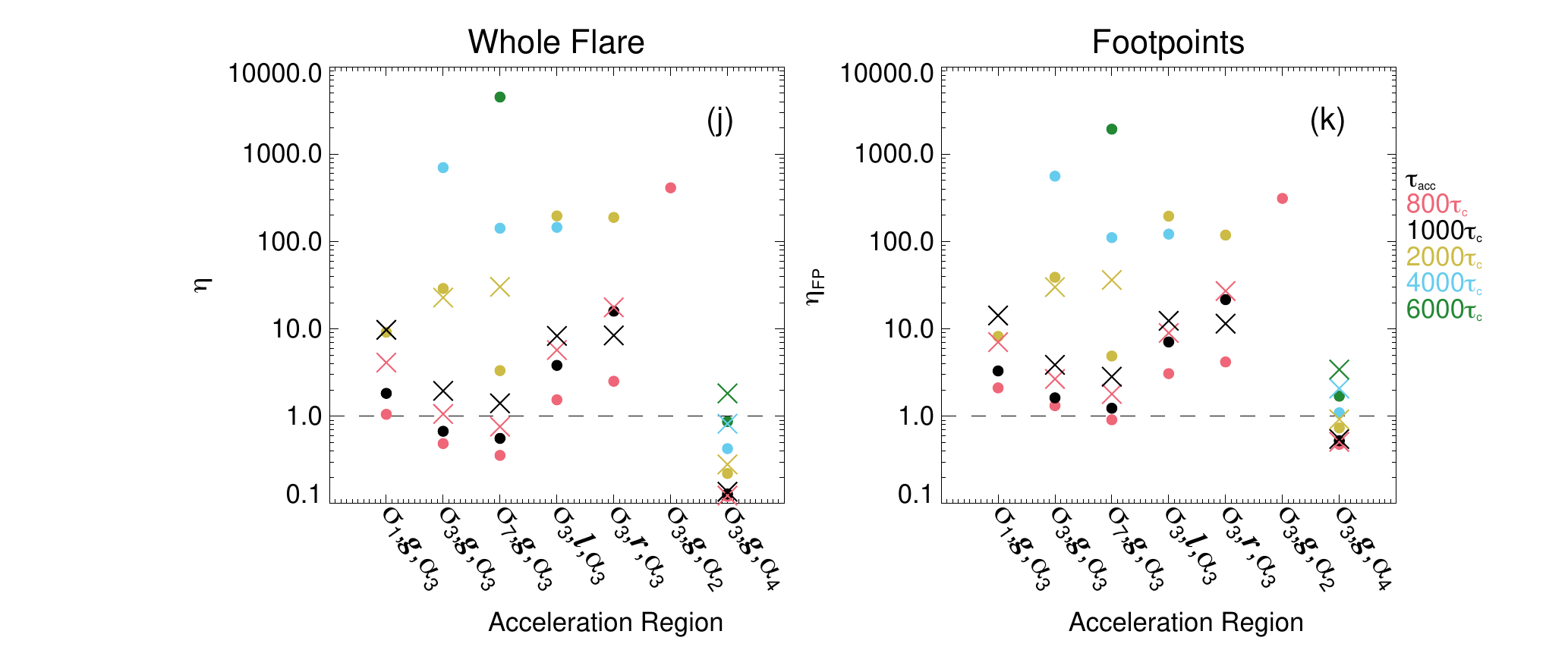}
    \caption{Panels a-f: Example spatial distributions $nVF$ [electrons s$^{-1}$ cm$^{-2}$ keV$^{-1}$] versus $|z|$ for different energy bins, for the control simulation, at (a and d) $\uptau_{acc} = 800\uptau_c$, (b and e) $2000\uptau_c$, and (c and f) $6000\uptau_c$, without (a-c) and with (d-f) the inclusion of shorter timescale turbulent scattering. Panels g-i: $\eta = (nVF(E=15-20\;{\rm keV}))/(nVF(E=50-100\;{\rm keV}))$ versus $\uptau_{acc}$ for different (g) spatial extents, (h) spatial functions, and (i) velocity dependencies. Crosses and circles show simulations with and without the inclusion of shorter timescale turbulent scattering. Each shaded region indicates the possible range of $\eta$ when considering the addition of a background thermal component (up to a maximum value of EM$=10^{49}$~cm$^{-3}$). Panels j and k: $\eta$ and $\eta_{\rm FP}= (nVF(E=20-30\;{\rm keV}))/(nVF(E=50-100\;{\rm keV}))$ versus $\uptau_{acc}$, for all simulations, respectively.}
    \label{fig:6}
\end{figure*}
Figures \ref{fig:6}a-\ref{fig:6}f plot angle-integrated $nVF$ versus $z$ for the control simulation using $\uptau_{acc} = 800\uptau_c$ (\ref{fig:6}a and \ref{fig:6}d), $2000\uptau_c$ (\ref{fig:6}b and \ref{fig:6}e) and $6000\uptau_c$ (\ref{fig:6}c and \ref{fig:6}f), with (\ref{fig:6}d -- \ref{fig:6}f) and without (\ref{fig:6}a -- \ref{fig:6}c) the inclusion of shorter timescale turbulent scattering, using a $1\arcsec$ spatial binning. 
In this figure, $z=0\arcsec$ indicates the loop apex with the chromospheric boundary at $\pm20\arcsec$.

The spatial distribution can change drastically for different energy bins.
For all spatial distributions and acceleration timescales, electrons with energy $E = 3-6$ keV have the highest electron flux in the coronal looptop, but $nVF$ for this energy range is significantly smaller in the footpoints. As electron energy increases, $nVF$ in the looptop decreases. 
Regardless of acceleration timescale, the two highest energy bins ($E = 25-50$ keV, $50-100$ keV) do not show a peak in the coronal looptop. This is consistent with observations where coronal X-ray sources are usually observed at $\approx 10-30$~keV. 


At high acceleration timescales, nVF at both the chromospheric boundary ($|z| = 20''$) and the loop apex ($z = 0''$) is dominated by low energy electrons. 
At lower acceleration timescales (e.g. $\uptau_{acc} = 800\uptau_c$, Figure \ref{fig:6}a), the footpoint emission generally increases. However the energy of the electrons at the footpoints appears to depend on the spatial function used. For $\rm{g},\alpha_{3}$ simulations (i.e, $\sigma_{1},\rm{g},\alpha_{3}$, $\, \sigma_{3},\rm{g},\alpha_{3}$, and $\sigma_{7},\rm{g},\alpha_{3}$) the footpoint emission is comprised of all energy bins, and as the spatial extent increases, higher energy electrons become more prominent. In comparison, simulations with  $\textit{l}$ and $\textit{r}$ have significantly less high energy particles in the footpoints. Instead, even at a low acceleration timescale, the $6-12$ keV emission is still the most prominent. 

The acceleration timescale at which high energy emission ($ E = 50 - 100$ keV) dominates over low energy emission ($ E = 15-20$ keV) can be seen in the Figures \ref{fig:6}g-\ref{fig:6}i which shows $\eta$ for different acceleration timescales, where 
\begin{equation}
    \eta = \frac{nVF(E=15-20 \text{ keV})}{nVF(E=50-100 \text{ keV})} \, . 
\end{equation}
The dashed line is $\eta = 1$. Simulations with and without turbulent scattering are shown by crosses and circles, respectively. The addition of a background thermal component is represented by the shaded rectangles, creating a range of values for $\eta$.
Several data points are  missing, particularly at high acceleration timescales, as there are no electrons with energy $50-100$ keV for that simulation.

Let us now consider simulations without turbulent scattering and no background thermal component. For all acceleration regions, $\eta$ increases over several orders of magnitude as acceleration timescale increases from $\uptau_{acc} = 800\uptau_c$ to $\uptau_{acc} = 6000\uptau_c$.
When $\uptau_{acc}\geq 1000\uptau_c$, low energy electrons dominate for all simulated regions (except for $\alpha_4$). Whereas when $\uptau_{acc} = 800\uptau_c$ acceleration regions which have large footpoint emission (i.e, $\sigma_3,\rm{g},\alpha_3$, $\, \sigma_7,\rm{g},\alpha_3$, and  $\sigma_{3},\rm{g},\alpha_4$) also have values of $\eta < 1$ , as high energy electrons dominate. 

The greater the spatial extent or the velocity dependence the smaller $\eta$ for a given acceleration timescale, such that $\eta[\sigma_1 > \sigma_3 > \sigma_7]$ and $\eta[\alpha_2 >\alpha_3 > \alpha_4 ]$. However, the separation of the velocity dependencies is much clearer than the spatial extents. 
Once again, $\alpha_{2}$ and $\alpha_{4}$ show two extremes of $\eta$, where $\alpha_{4}$ is the lower boundary for a given acceleration timescale. This is clearly shown in Figure \ref{fig:6}j which shows how $\eta$ changes for all spatial functions and acceleration timescales.  Unfortunately, simulations using $\alpha_{2}$ only have electrons with energy greater than 50~keV for $\uptau_{acc} = 800\uptau_c$.  For this timescale $\alpha_{2}$ creates an upper boundary for $\eta$. 
Using this diagnostic, it is difficult to separate the spatial functions $\textit{l}$, $\textit{r}$, and $\rm{g}$. 

The addition of short timescale turbulent scattering generally increases $\eta$ by up to an order of magnitude. The greater the acceleration timescale, the smaller the change in $\eta$ due to shorter timescale turbulent scattering. The addition of a background thermal component increases $\eta$ by several orders of magnitude, removing most trends in the graphs. The only property of the acceleration region that could still be easily determined is the velocity dependence.

Similar to the integrated spatial distribution, in the chromosphere ($|z| \geq 20 \arcsec$) we study $\eta_{\rm FP}$ where 
\begin{equation}
    \eta_{\rm FP} = \frac{nVF(E=20-30 \;{\rm keV})}{nVF(E=50-100 \;{\rm keV})} \, . 
\end{equation}
Figures \ref{fig:6}j and \ref{fig:6}k show $\eta$ and  $\eta_{\rm FP}$ for each simulation, respectively. Acceleration timescales are shown in different colors, indicated in the legend. 
Compared to $\eta$ for the entire spatial distribution, for $\eta_{\rm FP}$ the lower energy bin has been increased to better suit the chromosphere. 
When a background thermal component is not included in the corona, both $\eta$ and $\eta_{\rm FP}$ are almost identical. The differences between $\eta$ and $\eta_{\rm FP}$ are centred around $\eta = 1$. When $\eta > 1$, $\eta_{\rm FP}$ may be slightly smaller than $\eta$. Whereas when $\eta < 1$, $\eta_{\rm FP} \gtrsim \eta$. However these differences are always less than an order of magnitude. 

As with $\eta$, from $\eta_{\rm FP}$ the velocity dependence and spatial extents may be identified. However unlike $\eta$, $\eta_{\rm FP}$ does not suffer from additional background thermal effects and thus, is a useful diagnostic for studying acceleration properties.

\begin{figure}
    \centering
    \includegraphics[scale = 0.8]{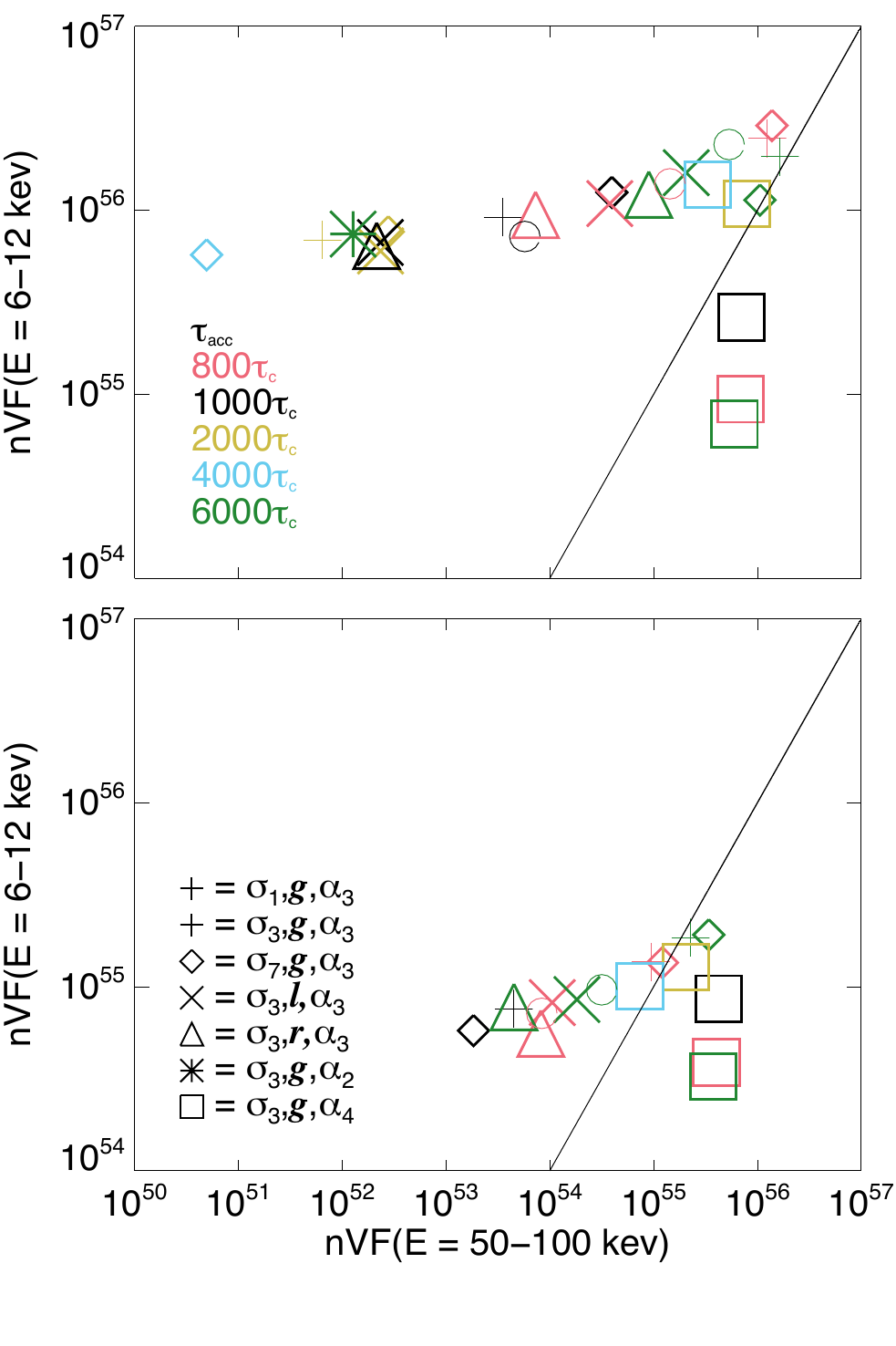}
    \caption{$nVF(E = 6-12$ keV) versus $nVF(E = 50-100$ keV) [electrons s$^{-1}$ cm$^{-2}$ keV$^{-1}$]. Simulations without (top) and with (bottom) the inclusion of shorter timescale turbulent scattering are displayed. Different acceleration timescales $\uptau_{acc}$ are shown using different colors with each acceleration region depicted by a different symbol.}
    \label{fig:7}
\end{figure}

Figure \ref{fig:7} shows $nVF(E = 6-12$ keV) against $nVF(E = 50-100$ keV) for each acceleration timescale, indicated by a different color, and each simulation, shown by a different symbol. Simulations `without' turbulent scattering are shown on the top and simulations with turbulent scattering are shown on the bottom. The black line is the identity line, when $nVF(E = 6-12$ keV) $= nVF(E = 50-100$ keV).

To the left of the identity line, for all simulations without turbulent scattering (except $\sigma_3,\rm{g},\alpha_{4}$), as acceleration timescale increases, $nVF(E = 50-100$ keV) increases over several orders of magnitude. In comparison, $nVF(E = 6-12$ keV) remains fairly consistent, increasing by just one order of magnitude across all simulations. There is a slight increase in $nVF(E = 6-12$ keV) which peaks to the left of the identity line. Unlike the other simulations, for simulations with $\alpha_{4}$, $nVF(E = 50-100$ keV) remains fairly constant, and $nVF(E = 6-12$ keV) decreases over an order of magnitude as acceleration timescale increases. 

Similar trends can be seen in simulations with turbulent scattering, except the range of values of $nVF(E = 50-100$ keV) is smaller, increasing over three orders of magnitude instead of six. Generally, simulations which include shorter timescale turbulent scattering have a lower $nVF(E = 50-100$ keV). 

For simulations without turbulent scattering, the only data to the right of the identity lines comes from simulations which use $\alpha_{4}$ for $800\uptau_c < \uptau_{acc} < 2000\uptau_c$. These data points represent the simulations in which high energy emission ($E = 50-100 $~ keV) dominates over low energy emission in the footpoints ($E = 6-12 $~keV). 
 
Overall, $\eta$ can be used to determine the velocity dependence. If the background thermal component is small the spatial extent and spatial function may also be determined. Furthermore,  plotting $nVF(E = 6-12$~keV) against $nVF(E = 50-100$~keV), again for a small thermal component, may indicate if short timescale turbulent scattering is present.

\subsubsection{Footpoint height versus energy}
\label{sect:footpoints}
In a collisional thick-target model, the depth electrons travel into the chromosphere changes with electron energy, with higher energy electrons moving deeper into the chromosphere.

\begin{figure*}[t!]
    \centering
    \includegraphics[scale = 0.85]{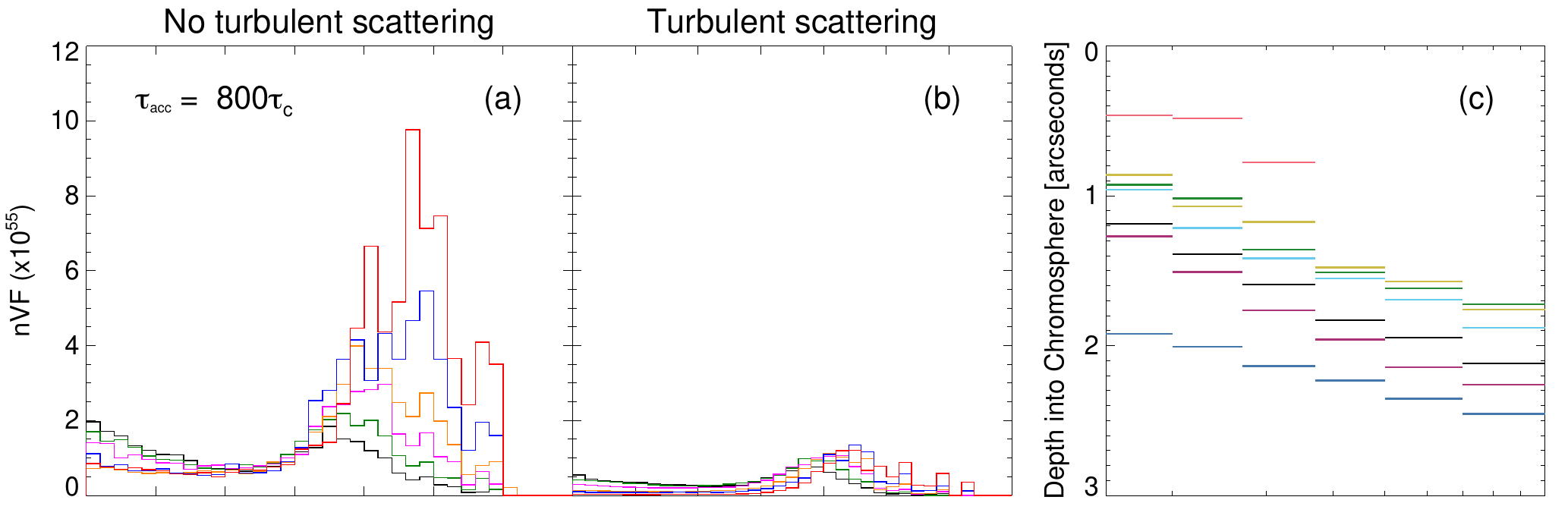}
    \includegraphics[scale = 0.85]{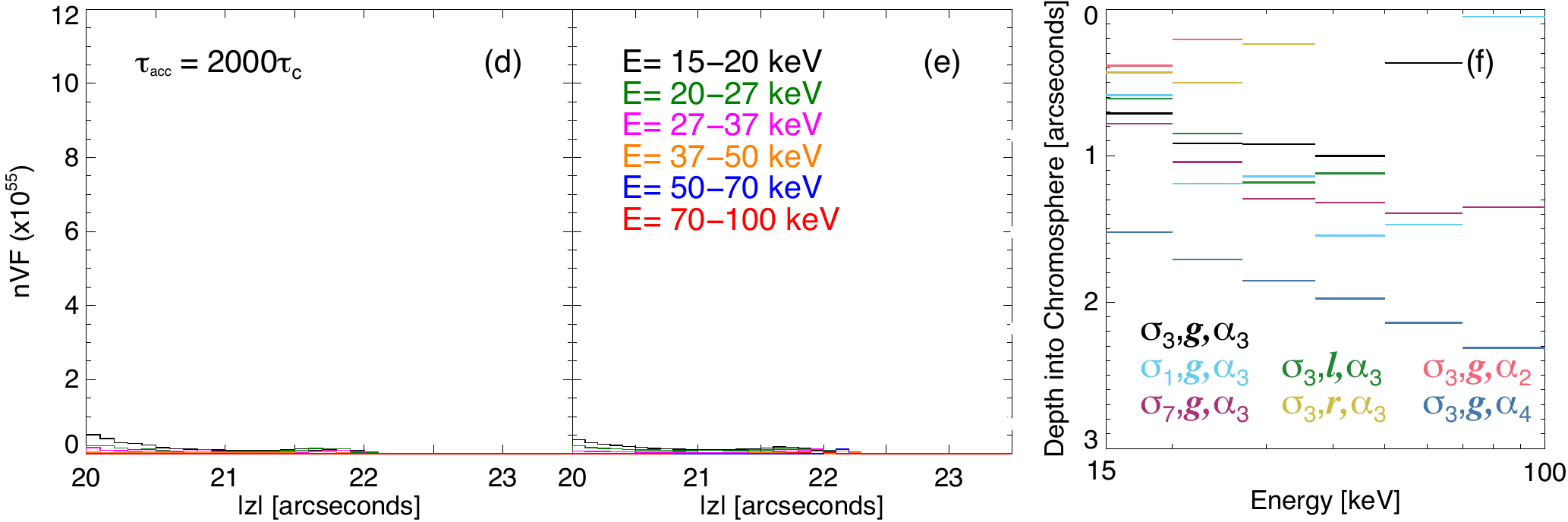}
    \caption{Panels (a), (b), (d), and (e): Spatial distributions ($nVF$ versus $|z|$) in the chromosphere for different energy bins for the control simulation ($\sigma_3,\rm{g},\alpha_{3}$, without and with the inclusion of shorter timescale turbulent scattering). Acceleration timescales $800\uptau_c$ and $2000\uptau_c$ are shown in (a)-(b) and (d)-(e), respectively, and $z=20\arcsec$ represents the top of the chromosphere. Panels (c) and (f): The mean depth electrons travelled into the chromosphere versus energy for simulations without turbulent scattering for an acceleration timescale of $800\uptau_c$ and $2000\uptau_c$, respectively. Here, $0\arcsec$ represents the top of the chromosphere.}
    \label{fig:8}
\end{figure*}

Figure \ref{fig:8} shows emission in the chromosphere. Figures \ref{fig:8}a and \ref{fig:8}d plot $nVF$ versus $|z|$ in the chromosphere using a $0.1\arcsec$ spatial resolution with no turbulent scattering, whilst Figures \ref{fig:8}b and \ref{fig:8}e show the spatial distribution with short timescale turbulent scattering. An acceleration timescale of $800\uptau_c$ (Figures \ref{fig:8}a-\ref{fig:8}c) and an acceleration timescale of $2000\uptau_c$ (Figures \ref{fig:8}d-\ref{fig:8}f) are both shown.
The colors show different electron energy bins, given in the legend. 

\begin{figure*}
    \centering
    \includegraphics[scale = 0.85]{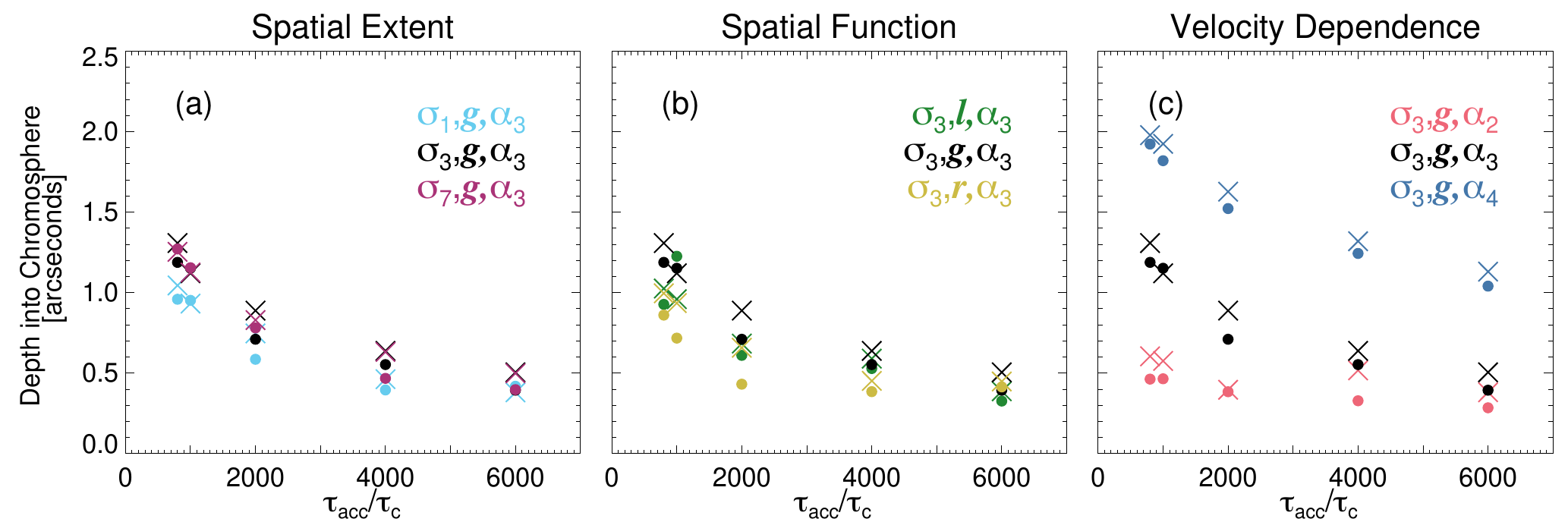}
    \caption{Average depth of electrons in the chromosphere between $15$ keV $\leq E < 100$ keV, versus acceleration timescale $\uptau_{acc}$ for different (a) spatial extents, (b) spatial functions, and (c) velocity dependencies. Crosses and circles show simulations with and without including shorter timescale turbulent scattering, respectively.}
    \label{fig:9}
\end{figure*}
Consider the simulations without short timescale (`no') turbulent scattering (Figures \ref{fig:8}a and \ref{fig:8}d). The spatial distribution in the chromosphere changes drastically depending on the amount of acceleration in the loop. 
When there is little acceleration (e.g., $\uptau_{acc} = 6000\uptau_c$) all energy bins show a peak in $nVF$ at the chromospheric boundary and $nVF$ decreases with chromospheric depth.
Alternatively, when there is efficient acceleration ($\uptau_{acc} = 800\uptau_c$) there is an initial peak in $nVF$ at the chromospheric boundary, where lower energy electrons are dominant. Then, we also see a secondary peak at $\approx 2\arcsec$ into the chromosphere, where higher energy electrons are dominant.

When we do account for shorter timescale turbulent scattering (Figures \ref{fig:8}b and \ref{fig:8}e) the chromospheric emission is reduced across all energy bins. Furthermore, the secondary peak in $nVF$ seen at small $\uptau_{\rm acc}$ almost fully disappears in the two simulations shown in Figure \ref{fig:8}.

Figures \ref{fig:8}c and \ref{fig:8}f show the average electron depth in the chromosphere versus electron energy for $\uptau_{\rm acc}=800\uptau_c$ and $\uptau_{\rm acc}=2000\uptau_c$, respectively. For a comparison with data, here the average depth is calculated using the first moment of $nVF$ versus $z$ for each energy bin\footnote{We use a mean value of the chromospheric depth since it is a better comparison with similar published observational results i.e., \citet{2010ApJ...717..250K} where the centroid positions of HXR footpoint sources are determined to sub-arcsecond accuracy.}. Here, a depth of $0\arcsec$ corresponds to the top of the chromosphere. All simulated acceleration regions without shorter timescale turbulent scattering are shown in different colors, given in the legend. 

Changing the velocity dependence creates upper and lower boundaries of electron depth. 
For any energy, electrons in $\alpha_2$ simulations appear at locations closer to the chromospheric boundary ($0\arcsec$) and for $\alpha_4$, electrons are located deeper in the chromosphere. Between these two boundaries are the other simulations (using $\alpha_3$).
As $\uptau_{acc}$ decreases, electrons of a give energy are located deeper in the chromosphere.
Thus, the initial depth of the chromospheric emission (given by low energy electrons) may help indicate the acceleration timescale and velocity dependence. 
At low acceleration timescales of $\uptau_{acc} = 800\uptau_c$, the electron depth into the chromosphere increases linearly with energy. 
For $\sigma_3,\rm{g},\alpha_4$, we see this trend for both acceleration timescales shown. However, the gradient of depth with energy increases as $\uptau_{acc}$ increases. This comparison cannot be made for other simulated acceleration regions, as at high acceleration timescales the distributions are more scattered; this may be due to a lack of electrons at higher energies.
The scattered distribution at higher acceleration timescales makes it difficult to separate the different spatial extents, which is possible at $\uptau_{acc} = 800\uptau_c$.
Such that, electron depth into the chromosphere for $\sigma_1 < \sigma_3 < \sigma_7$, for a given energy. It is difficult to distinguish between the $\textit{l}$ and $\textit{r}$ spatial functions. Once again, these functions may be better compared to a $\rm{g}$ distribution of a smaller spatial extent, i.e., $\sigma_1,\rm{g},\alpha_3$. 

Figure \ref{fig:9} shows the average depth for electrons of energy $15$ keV $\leq E < 100$ keV, versus acceleration timescale. As acceleration timescale increases, depth decreases, as expected. Although chromospheric $nVF$ is lower for simulations with shorter timescale turbulent scattering, the average depth for all electron energies between $15$ keV $\leq E < 100$ keV, is approximately the same, as seen in Figure \ref{fig:9}. 

The velocity dependence changes depth versus acceleration timescale such that we see greater chromospheric depths for larger $\alpha$. The greater the spatial extent, the greater the depth. However, it is difficult to distinguish between $\sigma_1$ and $\sigma_3$. The different spatial functions cannot be identified easily either. 

Overall, studying the emission in the chromosphere may help to constrain the acceleration timescale by studying (1) the gradient of depth versus energy and (2) the depth of low energy electrons ($E \lesssim 15$ keV) in the chromosphere. This diagnostic is useful for limb flares, such as that studied in \citet{2010ApJ...717..250K}. If the acceleration timescale is low, depth vs energy may also indicate the spatial extent of the acceleration region. Alternatively, if the acceleration timescale is high, the average depth versus acceleration timescale may constrain the velocity dependence.

\subsubsection{Coronal looptop}

\begin{figure*}[t!]
    \centering
    \includegraphics[scale = 0.85 ]{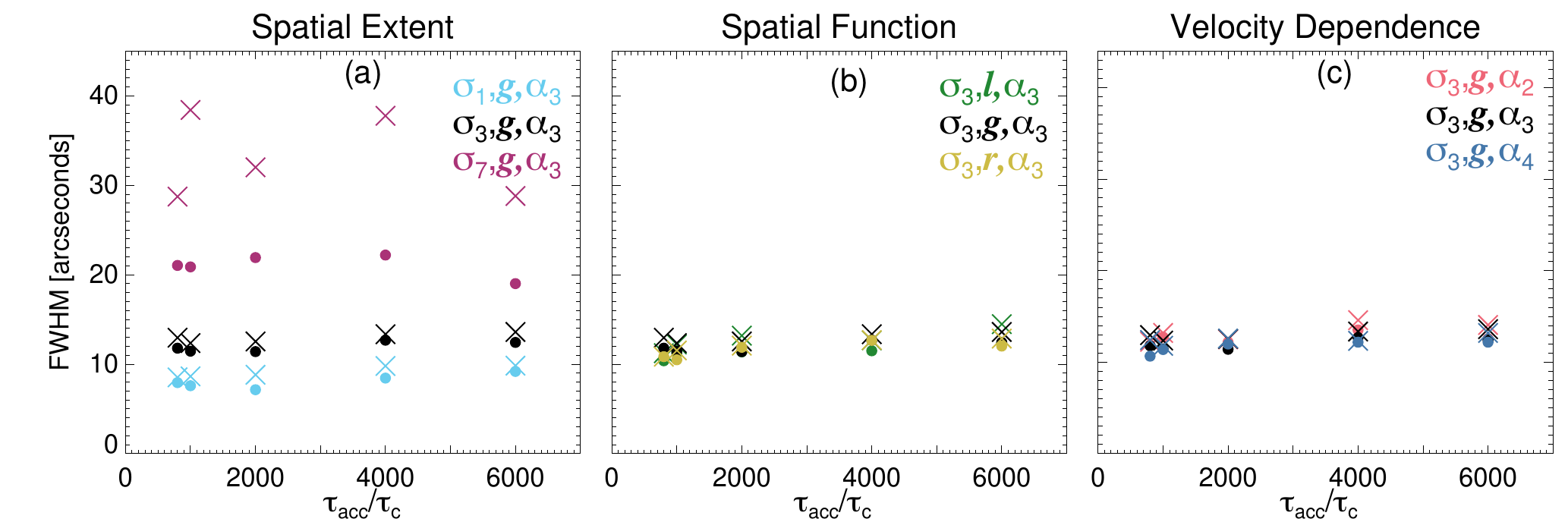}
    \caption{Coronal looptop FWHM versus $\uptau_{acc}$ for different (a) spatial extents, (b) spatial functions, and (c) velocity dependencies. Crosses and circles denote simulations with and without shorter timescale turbulent scattering.}
    \label{fig:10}
\end{figure*}

The spatial distribution in the coronal loop, $-20\arcsec <z < 20\arcsec$ (see the middle panel of Figure \ref{fig:2}) can be approximated with a Gaussian distribution in the majority of cases. For each acceleration region, the full width at half maximum (FWHM) [arcseconds] was determined for electrons with energy $10$ keV $< E< 15$ keV\footnote{The energy range is chosen to allow comparison with observational results, e.g., \cite{jeffrey2015high}, where the FWHM of a coronal looptop source is determined from X-ray data.} using: 
\begin{equation}
    \text{FWHM} = 2\sqrt{2\text{ln}2} \,  \Delta z \, ,
\end{equation}
where $\Delta z$ [arcseconds] is the standard deviation of the Gaussian distribution. Figure \ref{fig:10} shows the FWHM against $\uptau_{\rm acc}$ for each acceleration region studied. 
Some spatial distributions (e.g., $\sigma_1,\rm{g},\alpha_3$ at $800\uptau_{c}$) may be better fitted with a kappa distribution. However, due to the current instrumental resolution and the indirect techniques employed with X-ray data, a Gaussian distribution is suitable in determining the coronal source width (FWHM). 

For all simulations the measured FWHM is larger than the extent of the acceleration region (i.e., $2\sqrt{2\text{ln}2}\sigma$ for a Gaussian region). For example $\Delta z/\sigma=$ $3.4$, $1.6$, and $1.3$, for $\sigma_1$, $\sigma_3$, and $\sigma_7$, respectively. Thus, this ratio decreases as spatial extent increases. 

The FWHM is fairly constant as $\uptau_{acc}$ increases. Changing the velocity dependence and the spatial function does not have any clear impact on the FWHM. In comparison, the spatial extent of the acceleration region greatly affects the looptop FWHM (Figure \ref{fig:10}a), where FWHM[$\sigma_{7} >\sigma_{3} >\sigma_{1}$]. Thus, the FWHM of the coronal source is extremely useful for constraining the spatial extent of the acceleration region. Again, any coronal diagnostic can be affected by the presence of a separate background thermal component. Making the assumption that the majority of the background heating occurs in an identical location to the region of acceleration, this source would only increase the peak intensity of the coronal loop emission whilst the FWHM is not affected. 

The addition of shorter timescale turbulent scattering slightly increases the FWHM by $\sim 2\arcsec$ for each simulation. The exception to this is $\sigma_7,\rm{g},\alpha_3$ where the FWHM increases by up to $\sim15\arcsec$. 

\section{Discussion and Summary}\label{sec:summary}

This study has focused on an extended turbulent acceleration region and how changing the spatial extent, the spatial distribution, and the velocity dependence of the acceleration region affects the electron distribution for various acceleration timescales, with the purpose of finding useful spectral and imaging diagnostics that can constrain the acceleration properties from X-ray data alone. 

For several diagnostics ($\delta_{nVF}$, $\varphi$, $\eta$, depth into chromosphere) setting the velocity dependence to $\alpha = {2}$ and $\alpha = 4 $ provides approximate upper and lower acceleration boundaries for the bulk of acceleration times studied. When $\alpha = 2$ the resulting electron distribution is mainly thermal and concentrated in the coronal loop. These electrons do not travel far into the chromosphere. Alternatively, $\alpha =4$ rapidly accelerates electrons in the corona and electrons propagate deep into the chromosphere. As a result, the energy spectrum for $\alpha=4$ is very flat for all acceleration timescales and the $nVF$ spatial distribution contains large peaks in the  chromosphere and is significantly flatter in the corona. 
Between these two extremes are simulations when $\alpha = 3$. 

The difference in spectral index for the spatially resolved spectra in the looptop and footpoint regions is suggestive of electron confinement, $\delta_{nVF}^{LT}-\delta_{nVF}^{FP} < 2$. Here, this difference occurs for low acceleration efficiency (i.e., large $\uptau_{acc}$), and no additional trapping mechanism is required. Not usually studied, the ratio between the spectral index at the looptop and footpoints ($\delta_{nVF}^{LT}/\delta_{nVF}^{FP}$) exponentially decreases as the acceleration timescale $\uptau_{acc}$ increases. 
The spatial distribution of the emitting electrons drastically changes with $\uptau_{acc}$. For large $\uptau_{acc}$, emission occurs almost exclusively in the coronal looptop source. However, at small $\uptau_{acc}$, emission becomes significant in the coronal footpoints.

Combining diagnostics helps to determine the properties of the acceleration region. If both X-ray spectral and imaging diagnostics are available, we suggest following this method to constrain the acceleration region properties: 
\begin{enumerate}
    \item \textit{Spatial Extent:} this can be estimated from the FWHM of the coronal loop. As discussed, the measured FWHM of the loop is always larger than that of the actual acceleration region. If the observed flare has a small background thermal component then $\varphi$,  the ratio of looptop to footpoint $nVF$, may be used to determine the spatial extent. Alternatively, if the background thermal component is large, the depth lower energy ($< 20$ keV) electrons reach in the chromosphere and the gradient of electron depth with energy, may provide some insight on the spatial extent. However, this can only be applied at low acceleration timescales.  
    
    \item \textit{Velocity Dependence:} the spectral index can be used to determine the velocity dependence, where a larger velocity dependence has a smaller spectral index. Further to this, if the observed flare has a small background thermal component, $\varphi$ and $\eta$, the ratio of $nVF$ at specific energy ranges, may also be used to determine the velocity dependence. To further investigate the velocity dependence, determine electron depth into the chromosphere for different energies if possible (i.e., limb flares).

    \item \textit{Spatial Function}: Identifying the spatial function is the most difficult. 
    If the background thermal component of the coronal loop is small, $\varphi$ may be used to determine the spatial function of the acceleration region. If the background thermal component is large, then begin by examining at the spectral index. A linear and random function resulted in a spectral index which was larger than that of a Gaussian function. Yet, the FWHM of the coronal looptop is similar for linear, random and Gaussian functions. Observing this may indicate a non-Gaussian spatial distribution. Distinguishing between non-Gaussian distributions is harder, but again, this may be achieved with $\varphi$ if the background thermal component is small. 
    Further, if EUV spectral data is available, then it might be possible to constrain the spatial distribution by combining the data.
    
    \item \textit{Acceleration timescale}: After determining the spatial extent, spatial function and velocity dependence, the acceleration timescale should be constrained. Information regarding the acceleration timescale may be determined from studying the ratio of spectral index in the looptop to footpoints ($\delta_{nVF}^{LT}/\delta_{nVF}^{FP}$). If the ratio $> 1$ then the acceleration timescale is small. Furthermore, studying how electron depth into the chromosphere changes with energy may help to determine $\uptau_{acc}$. In addition, the deeper the low energy ($\approx 15-20$ keV) emission into the chromosphere, the shorter the acceleration timescale. 
    
    \item \textit{Level of turbulent scattering}: The presence of shorter-timescale turbulent scattering leads to spectral steepening. Alternatively, $\varphi$ appears to decrease with the presence of shorter-timescale turbulent scattering. If the acceleration timescale is small, the emission in the chromospheric footpoints may also indicate if there is short-timescale turbulent scattering present, as the emission drastically decreases compared to when there is no turbulent scattering. 
\end{enumerate}

The simulations shown here are obviously a small set showing the applicability of different X-ray spectral and imaging diagnostics. We cannot simulate all plasma properties or every possible transport mechanism in the corona. Next, we will apply our diagnostics to suitable flares using archived RHESSI data and new SolO/STIX data, and simulations will be fine-tuned to match the electron and plasma properties of an individual flare, and the position of the spacecraft in the heliosphere (any parameters with units of arcsecond here are calculated at a distance of 1~AU).
\begin{figure}[t!]
    \centering
    \includegraphics[scale = 0.9]{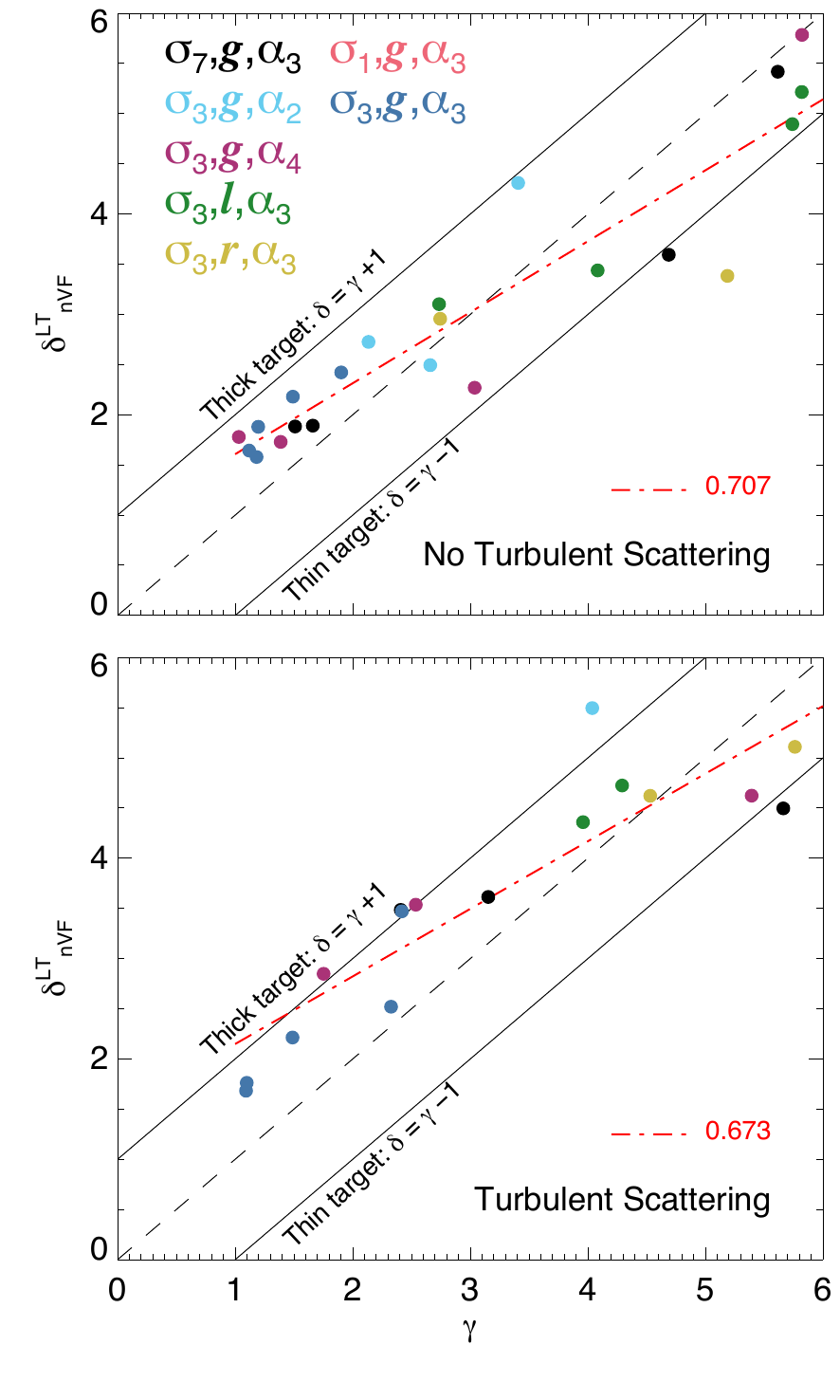}
    \caption{$\delta_{nVF}^{LT}\approx \delta_{\rm 1AU}$ versus $\gamma = \delta_{nVF} + 1$ for all modelled acceleration regions, and representing different flare observations where we see different populations of electrons at the Sun (producing HXRs) and in-situ in the heliosphere. The gradient of the line of best fit (pink dashed-dot) for simulations without turbulent scattering (top) and with short timescale turbulent scattering (bottom) is 0.707 and 0.673, respectively, consistent with an approximate gradient of 0.70 in \cite{2007ApJ...663L.109K} for prompt events. The solid lines on each graph represent the thick and thin target models.}
    \label{fig:11}
\end{figure}

Lastly, \cite{2007ApJ...663L.109K} compared the HXR photon spectral index $\gamma$, measured remotely at the Sun, with the spectral index of related flare-accelerated electrons measured in-situ at 1AU, $\delta_{1\text{AU}}$. 
The results of our combined acceleration and transport modelling are consistent with \cite{2007ApJ...663L.109K}, if we plot $\delta_{nVF}^{LT}\approx\delta_{1\text{AU}}$ (Figure \ref{fig:11}) against $\gamma = \delta_{nVF} + 1$, where we have assumed that $\delta_{nVF}^{LT}$ in the corona is identical to $\delta_{1\text{AU}}$, and that both sets of electrons are produced in the same acceleration region. This suggests that there is no need to employ secondary acceleration models when observations are not fully consistent with either a thin or thick target model.
We are now investigating this in a more considered manner using electron modelling in the heliosphere and a comparison with HXR-producing electrons at the Sun. The use of in-situ data, when available, alongside the suggested diagnostics will help to further constrain the properties of the solar flare acceleration environment.

\section*{acknowledgments}
MS gratefully acknowledges financial support from a Northumbria University Research Development Fund (RDF) studentship. NLSJ gratefully acknowledges the current financial support from the Science and Technology Facilities Council (STFC) Grant ST/V000764/1. JAM gratefully acknowledges financial support from  STFC grant ST/T000384/1. The authors acknowledge IDL support provided by STFC. MS and NLSJ are supported by an international team grant \href{https://teams.issibern.ch/solarflarexray/team/}{“Measuring Solar Flare HXR Directivity using Stereoscopic Observations with SolO/STIX and X-ray Instrumentation at Earth}” from the International Space Sciences Institute (ISSI) Bern, Switzerland. 

\bibliography{main_text}{}
\bibliographystyle{aasjournal}

\end{document}